\documentclass[pre,groupedaddress,twocolumn,floatfix]{revtex4-2}

\usepackage{microtype}
\usepackage[utf8]{inputenx}
\usepackage{physics}                            
\usepackage{bm}									
\usepackage{enumerate}							
\usepackage{amsthm}								
\usepackage{comment}							
\usepackage[caption=false]{subfig}
\usepackage{amssymb}							
\usepackage{t1enc}								
\usepackage{mathtools}							
\usepackage{tensor}								
\usepackage{wrapfig}							
\usepackage{listings}							
\usepackage{xcolor}								
\usepackage{hyperref}
\hypersetup{colorlinks=true}
\usepackage[capitalize]{cleveref}               
\usepackage{graphicx,color}
\usepackage{amsfonts,amsmath}
\usepackage{xstring}                            
\usepackage{lipsum}

\newcommand{\cor}[1]{\mathcal{#1}}									
\newcommand{\T}[1]{\text{#1}}										
\newcommand{\dslash}[1]{\frac{\dd[d]{#1}}{(2\pi)^d}}                
\def \z {^{(0)}}

\def \ss {^\T{(ss)}}
\def \sc {^\T{(sc)}}
\def \v {^\T{(v)}}
\newcommand{\eg}{e.g.}
\newcommand{\ie}{i.e.}
\newcommand{\n}{\nonumber}
\newcommand{\ccite}[1]{\IfSubStr{#1}{,}{Refs.~}{Ref.~}\cite{#1}}

\newcommand{\rev}[1]{#1}

\begin{document}
\title{Memory-induced oscillations of a driven particle in a dissipative correlated medium}
\author{Davide Venturelli}
\email{dventure@sissa.it}
\author{Andrea Gambassi}
\affiliation{SISSA --- International School for Advanced Studies and INFN, via Bonomea 265, 34136 Trieste, Italy}

\begin{abstract}
The overdamped dynamics of a particle is in general affected by its interaction with the surrounding medium, especially out of equilibrium, and when the latter develops spatial and temporal correlations.
Here we consider the case in which the medium is modeled by a scalar Gaussian field with relaxational dynamics, and the particle is dragged at constant velocity through the medium by a moving harmonic trap.
This mimics the setting of an active microrheology experiment conducted in a near-critical medium. 
When the particle is displaced from its average position in the nonequilibrium steady state, its subsequent relaxation is shown to feature damped oscillations. This is similar to what has been recently predicted and observed in viscoelastic fluids, but differs from what happens in the absence of driving or for an overdamped Markovian dynamics, in which cases oscillations cannot occur. 
We characterize these oscillating modes in terms of the parameters of the underlying mesoscopic model for the particle and the medium, confirming our analytical predictions via numerical simulations.
\end{abstract}

\maketitle

\section{Introduction}
\label{sec:intro}

Complex media with macroscopic relaxation timescales and spatial correlations are not expected to be a source of white noise for the stochastic dynamics of a particle immersed into them. 
In fact, the assumption of timescale separation, underlying the use of a simple Langevin equation~\cite{Langevin_1908} for describing the dynamics of the mesoscopic particle, is no longer valid when the particle motion and the evolution of the medium occur over comparable timescales~\cite{Dhont_1996}. 
Moreover, in the presence of an external driving force acting on the particle, the surrounding medium to which the particle is (weakly) coupled can no longer be assumed to remain in equilibrium, as in the case of the undriven Brownian motion. In fact, the medium is expected to react to the passage of the particle, and thus to be generically out of equilibrium.

In this respect, a notable example of such complex media is provided by viscoelastic fluids~\cite{Larson_1999}: their non-Newtonian behavior originates from the storage and dissipation of energy within their complex microstructure, which translate into a macroscopically long stress-relaxation time. 
Dragging a colloidal particle through such a fluid --- as it is typically done in active microrheology experiments~\cite{Squires_2005,Gazuz_2009,GomezSolano_2014,GomezSolano_2015,Jain_2021,Jain_2021_2step} --- drives the medium out of equilibrium. In turn, this affects the statistics of the particle position~\cite{Dhont_1996}. At a coarse-grained scale, the resulting particle dynamics is often described by an overdamped generalized Langevin equation (GLE~\cite{Mori_1965,Zwanzig_book}). 
In this equation, the effect of the interaction between the particle and the medium is encoded in a friction kernel $\cor{K}(t)$ acting on the particle velocity as $\int_{-\infty}^t \dd{u} \cor{K}(t-u) \dot{X}(u) = F(X,t) + \zeta(t)$, where $F$ includes the forces exerted on the particle at position $X$, while $\zeta$ is a 
colored Gaussian noise.  

Recently, it has been experimentally shown that viscoelasticity can give rise to oscillating modes in the overdamped motion of colloidal particles driven through the medium~\cite{Berner2018}. 
This is somewhat
unexpected and noteworthy, because oscillations (which typically occur in systems with underdamped dynamics) are strictly forbidden at equilibrium, as shown, e.g., in~\ccite{Berner2018}. 
Heuristically, one may note that integrating by parts the retarded friction in the GLE above formally renders a term $\int_{-\infty}^t \dd{u} \cor{M}(t-u) \ddot{X}(u)$, where $\cor{M}(t) \sim \int^t \dd{u} \cor{K}(u)$ can be readily interpreted (if positive) as a memory-induced \textit{inertia} \cite{Zwanzig_book}, which is generally absent from the description of Markovian overdamped systems.

Memory terms in the effective evolution equation of a particle actually appear quite naturally in many physical systems, after integrating 
the slow degrees of freedom out of the original, microscopic dynamics in which they are coupled to those describing the tracer particle \cite{mori_contraction_1980,morita_contraction_1980,teVrugt_2020}. 
For example, a minimal model for diffusion in a thermally fluctuating correlated medium can be 
formulated in terms of the joint overdamped dynamics of a particle and of a scalar Gaussian field $\phi(\vb{x},t)$, the latter being characterized by a
correlation length $\xi$ and a finite relaxation time \cite{demery2010, demery2010-2, demery2011, demerypath, Dean_2011, demery2013, Gross_2021,Venturelli_2022, wellGauss,Venturelli_2022_2parts,Venturelli_2022_confined}. 
If the coupling between the field and the particle is chosen to be linear, then the field can be integrated out exactly, resulting into an effective evolution equation for the particle. 
This equation provides insight on the connection between the emerging memory kernel and the features of the original microscopic model. 
The system described here may be viewed as a toy model for a colloidal particle in contact with a fluid medium in the vicinity of a critical point, such as a binary liquid mixture, which displays long-range spatial correlations and long relaxation times. In this specific example, the field $\phi$ represents the order parameter associated with the second-order phase transition, while hydrodynamic effects and other slow variables \rev{that} should be taken into account when describing real fluids \cite{halperin} are neglected for simplicity. 
In recent years, this kind of physical systems have been the subject of various experimental investigations \cite{Hertlein_2008,Gambassi_2009,Paladugu_2016,Ciliberto_2017,Magazzu_2019},
especially concerning the emergence of effective, critical Casimir forces mediated by the fluctuations of the medium \cite{Krech_book,Danchev_book,GambassiCCF}.
More generally, analogous spatio-temporal correlations also characterize the dynamics of such diverse physical systems as inclusions in lipid membranes \cite{reister_lateral_2005,reister-gottfried_diffusing_2010,camley_fluctuating_2014}, microemulsions \cite{Gompper_1994,Hennes_1996,Gonnella_1997}, or defects in ferromagnetic systems \cite{demery2010, demery2010-2, demery2011, demerypath, Dean_2011, demery2013}.

In this context, it is natural to ask whether the memory kernel in the effective evolution equation of the tracer particle, which originates from the spatio-temporal correlations of the field, may give rise to oscillating modes similar to those observed in viscoelastic fluids \cite{Berner2018}
--- which, instead, are primarily due to the mechanical response of the medium.
In order to address this question, we consider here the simple setting of a particle driven through the medium at a constant velocity $v$ by a moving harmonic potential, while being in contact with a scalar Gaussian field subject to an overdamped relaxational dynamics \cite{Tauber}.
(The driving considered here can be practically realized via optical tweezers \cite{jones2015optical}.)
We first integrate out the field degrees of freedom, thus obtaining an effective (non-linear) equation which describes the motion of the particle in the steady state reached by the system at long times. 
By linearizing this equation and inspecting the analytic structure of the field-induced memory kernel, we demonstrate that damped oscillations are indeed displayed by the particle during the relaxation \rev{that} occurs after it has been displaced from its steady-state position. These oscillations are confirmed via numerical simulations of the model.
The simplicity of our model allows us to study in detail how the interplay between the various timescales of the system 
dictates the emergence of the particle oscillations, and to determine their frequency and typical decay time.

The rest of the presentation is organized as follows. In Sec.~\ref{par:Model} we introduce the model and we characterize the steady state attained at long times by the particle in the moving trap. In Sec.~\ref{sec:noiseless} we analyze the 
relaxation of a particle initially displaced from its steady-state position, under the assumption that thermal fluctuations are negligible. The effect of these fluctuations is then assessed in Sec.~\ref{sec:thermal}, where we also compare our analytical predictions with numerical simulations. We finally summarize our findings and present our conclusions in Sec.~\ref{par:conclusion}.

\begin{figure}[t]
    \centering
    \includegraphics[width=0.9\columnwidth]{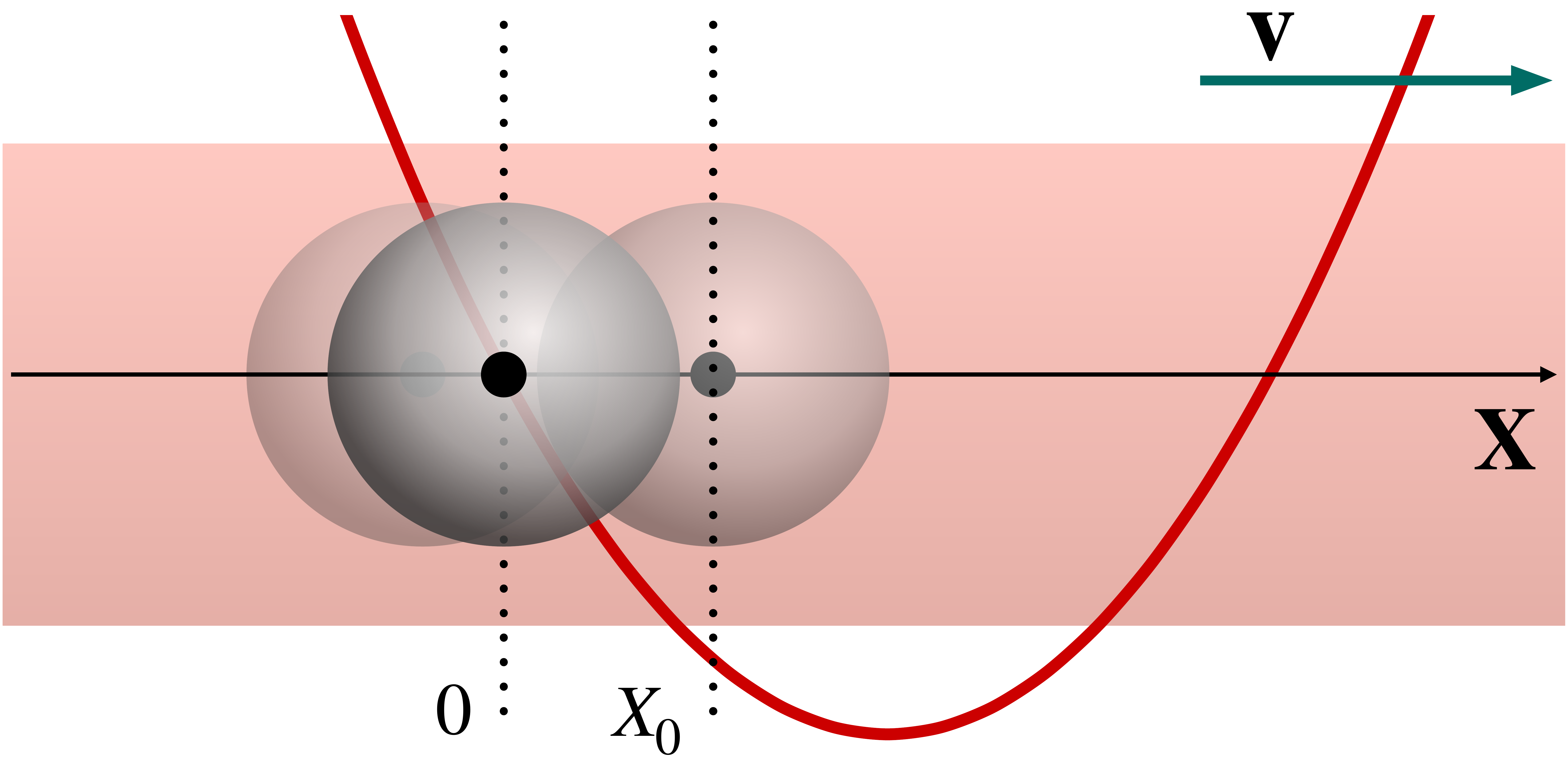}
    \caption{Schematic representation of the system under investigation. A particle is spatially confined by a harmonic potential 
    (realized, \eg, by optical tweezers), the center of which is dragged at constant velocity $\vb{v}$. The particle interacts with a thermally fluctuating order parameter $\phi$ (red background) according to the Hamiltonian in \cref{eq:hamiltonian}. The particle position $\vb{X}(t)$ is measured in a comoving frame of reference, chosen such that  $\expval{\vb{X}}=0$ in the steady state (see \cref{sec:noiseless} for details). At time $t=0$ the particle is suddenly displaced by a small amount $\vb{X}_0$ from its current position, and the ensuing relaxation is observed.
    }
    \label{fig:setup}
\end{figure}
%
%

\section{The model} 
\label{par:Model}
We consider a particle at position $\vb{Y}(t)\in\mathbb{R}^d$ in $d$ spatial dimensions, which is confined by means of a moving harmonic potential 
\begin{equation}
    \cor{U}(\vb{Y},t)=\frac{\kappa}{2}\left(\vb{Y} - \mathbf{v} t\right)^2
    \label{def:Uk}
\end{equation}
of stiffness $\kappa$, the center of which is dragged at a constant velocity $\vb{v}$.
The particle (solely described by the position $\vb{Y}$ of its center) is in contact with a correlated medium, which we model for simplicity as a fluctuating scalar order parameter field $\phi(\vb{x},t)\in \mathbb{R}$. 
The fluctuations of the latter are assumed to be characterized by the quadratic 
Hamiltonian \cite{halperin}
\begin{equation}
        \cor{H}_\phi= \int \dd[d]{\vb{x}}\left[ \frac{1}{2}(\nabla\phi)^2+\frac{1}{2}r\phi^2\right],
        \label{eq:gaussian_hamiltonian}
\end{equation}
where $\xi=r^{-1/2}\geq 0$ is the correlation length, which controls the spatial range of the field correlations at equilibrium, and diverges upon approaching the critical point $r=0$. 
The interaction between the particle and the field is chosen as \cite{wellGauss,Venturelli_2022,Venturelli_2022_2parts,heat} 
\begin{equation}
    \cor{H}_\T{int} = \rev{-\lambda} \int \dd[d]{\vb{x}} \phi(\vb{x})V(\vb{x}-\vb{Y}),
    \label{eq:Hint}
\end{equation}
so that the system composed by the particle and the field is described by the total Hamiltonian
\begin{equation}
    \mathcal{H}[\phi,\vb{Y}] = \cor{H}_\phi[\phi]  
    \rev{+}
    \cor{H}_\T{int}[\phi,\vb{Y}] +  \cor{U}(\vb{Y},t) .
    \label{eq:hamiltonian}
\end{equation} 
The system under investigation is schematically described in Fig.~\ref{fig:setup}.
The coupling in \cref{eq:Hint} is linear and translationally invariant,
while the interaction potential $V(\vb{x})$ models the shape of the particle. We choose $V(\vb{x})$ to be normalized so that its integral over all space is equal to one. With this normalization, the strength of the field-particle interaction is set only by the coupling constant $\lambda$. If $\lambda \, V(\vb{x})$ in Eq.~\eqref{eq:hamiltonian} is chosen to be positive, then field configurations are favored in which 
$\phi$ is locally enhanced, and therefore it assumes preferentially positive values in the vicinity of the particle. At the same time, the particle experiences an attractive force directed along the gradient of the field.
We make the assumption that the interaction potential $V(\vb{x})=\overline{V}(x/R)$ is isotropic and characterized by a single length scale, namely the ``radius'' $R$ of the particle. For example, we may choose an exponentially decaying potential
\begin{equation}
    V(\vb{x}) = \frac{1}{\Omega_d \Gamma_E(d)R^d} \exp( -\norm{\vb{x}}/R ),
    \label{eq:potential_peak}
\end{equation}
where $\Omega_d$ is the $d$-dimensional solid angle, and $\Gamma_E(z)$ is the Euler gamma function.

The physical dimensions $[\phi]$ and $[\lambda]$ of the field and the coupling, respectively, follow from the dimensional analysis of the Hamiltonian in Eq.~\eqref{eq:hamiltonian}. In units of energy $\cor{E}$ and length $\cor{L}$, they are given by $[\phi]= \cor{E}^{1/2} \cor{L}^{1-d/2} $ and $[\lambda]= \cor{E}^{1/2} \cor{L}^{d/2-1}$.
These expressions facilitate the dimensional analysis of the quantities introduced further below.

\subsection{Dynamics}

The dynamics of the particle is here described by the overdamped Langevin equation
\begin{align}
        \dot{\vb{Y}}&= -\nu \grad_{\vb Y} \cor{H}   + \bm{\xi} \label{eq:particle} \\
        &=-\nu \kappa(\vb{Y}-\vb{v}t) + \nu\lambda \int \dslash{q} i \vb{q} \phi_q V_{-q} e^{i\vb{q}\cdot \vb{Y}} + \bm{\xi}, \n
\end{align}
where $\nu$ is the mobility of the particle, $\phi_q = \int \dd{\vb{x}} \phi(\vb{x},t) \exp(-i \vb{q}\cdot \vb{x
}) $, 
and analogously $V_q$ is the Fourier transform of $V(\vb{x})$ \footnote{We normalize the delta distribution in Fourier space as $\int [\dd[d]{q}/(2\pi)^d] \delta^d(\vb{q})=1$.};
finally,
$\bm{\xi}(t)$ is a white Gaussian noise with zero mean and variance ($k_B\equiv 1$)
\begin{equation}
    \expval*{\xi_{i}(t) \xi_{j}(t') } = 2\nu T \delta_{ij}\delta(t-t'),
    \label{eq:part_noise}
\end{equation}
where $T$ is the temperature of the thermal bath (see below).
Similarly, we assume a purely relaxational dynamics for the field \cite{Tauber}, i.e., 
\begin{align}
        &\partial_t\phi(\vb{x},t)= -D (i \grad)^\alpha \fdv{\cor{H}}{\phi(\vb{x},t)} + \eta(\vb{x},t)  \label{eq:field} \\
        &= -D(i \grad)^\alpha \left[(r-\nabla^2)\phi(\vb{x},t)-\lambda V(\vb{x}-\vb{Y}(t)) \right] + \eta(\vb{x},t). \n
\end{align}
Here $\alpha=0$ for a non-conserved dynamics of the order parameter $\phi$, while $\alpha=2$ if $\phi$ is subject to a local conservation during the evolution. In this case, Eq.~\eqref{eq:field} can be cast in the form $\partial_t\phi(\vb{x},t)=-\div \vb{J}(\vb{x},t)$ with a suitable current $\vb{J}(\vb{x},t)$. The two choices of $\alpha$ correspond to model A ($\alpha=0$) and model B ($\alpha=2$) in the classification of Ref.~\cite{halperin}, in which we neglect the self-interaction term $\propto \phi^4$ (\ie, we consider the Gaussian approximation of these models). 
The particle and the field are assumed to be in contact with the same thermal bath at temperature $T$, so that $\eta(\vb{x},t)$ is also a  Gaussian white noise with zero mean and variance
\begin{equation}
    \expval*{\eta(\vb{x},t)\eta(\vb{x}',t')}= 2DT (i \grad)^\alpha \delta^d(\vb{x}-\vb{x}')\delta(t-t'),
    \label{eq:corr-eta-phi}
\end{equation}
where $D$ is the mobility of the field. In the absence of external dragging, \ie, for $v\equiv \norm{\vb{v}}=0$, the coupled dynamics of the field and of the particle satisfies detailed balance, and therefore the stationary state is described by the equilibrium canonical distribution $P_\T{eq}[\phi,\vb{Y}] \propto \exp(-\cor{H}[\phi,\vb{Y}]/T)$ \cite{Venturelli_2022}, with $\cor{H}$ given in Eq.~\eqref{eq:hamiltonian}.

In view of deriving the effective dynamics of the particle, it is convenient to write Eq.~\eqref{eq:field} in Fourier space as
\begin{align}
    &\dot{\phi}_q = -Dq^\alpha(q^2+r) \phi_q +\lambda D q^\alpha V_q e^{-i\vb{q}\cdot \vb{Y}}+ \eta_q ,
    \label{eq:fieldFourier} \\[2mm]
    &\expval*{\eta_q(t)\eta_{q'}(t')}= 2DTq^\alpha \delta^d(\vb{q}+\vb{q}')\delta(t-t').
    \label{eq:field_noise}
\end{align}
Upon setting $\lambda = 0$, \cref{eq:fieldFourier,eq:particle} reduce to a collection of non-interacting Ornstein-Uhlenbeck processes --- one for each of the $d$ components of the position of the particle, and one for each of the field modes $\phi_q$ (which form a continuum in the bulk, i.e., $\vb{q}\in \mathbb{R}^d$). 
These processes are characterized by the inverse relaxation timescales (see \cref{par:freefield})
\begin{align}
    \tau_\kappa^{-1} &= \gamma \equiv \nu \kappa , \label{eq:tau_kappa} \\ 
    \tau_\phi^{-1}(q) &= \alpha_q \equiv Dq^\alpha(q^2+r). \label{eq:tau_phi}
\end{align}
Accordingly, $\phi(\vb{x},t)$ is a medium which is correlated over both space and time, and in which the corresponding ranges are determined by $\xi$ and $\tau_\phi$, respectively.
In particular, the relaxation time $\tau_\phi(q\sim 0)$ of the long-wavelength modes of the field becomes arbitrarily long for model A dynamics at $r=0$. The same happens for model B with generic values of $r$, \ie, also off-criticality, due to the presence of the conservation law \cite{Tauber}.
These long-wavelength modes are always present in the bulk, while they are cut-off in a confined geometry such as that considered in Refs.~\cite{Gross_2021,Venturelli_2022_confined}.

\subsection{Steady state in the comoving frame}
\label{sec:steady state}

We start by measuring the position $\vb{Z}\equiv \vb{Y} -\vb{v} t$ of the particle in the frame of reference \rev{that} is comoving with the harmonic trap. 
In terms of the coordinate $\vb{Z}$, the equations of motion \eqref{eq:particle} and \eqref{eq:field} become 
\begin{align}
    &\dot{\vb{Z}} = -\vb{v} -\gamma \vb{Z} 
    \rev{-}\nu
    \grad_{\vb Z} \cor{H}_\T{int}[\varphi,\vb{Z}] + \bm{\xi} , \label{eq:EOM_frame_Z} \\[2mm]
    &(\partial_t- \vb{v}\cdot \grad) \varphi(\vb{x},t)  = -D (i \grad)^\alpha \fdv{\cor{H}[\varphi,\vb{Z}]}{\varphi(\vb{x},t)} + \eta(\vb{x},t) \n ,
\end{align}
where we introduced the translated field $\varphi(\vb{x},t)\equiv \phi(\vb{x}+\vb{v}t,t)$.
Note that $\cor{H}_\phi[\varphi]=\cor{H}_\phi[\phi]$ and $\cor{H}_\T{int}[\varphi,\vb{Z}]=\cor{H}_\T{int}[\phi,\vb{Y}]$, by translational invariance (which applies also to the white noises $\bm{\xi}$ and $\eta$). 
In Fourier space, these equations can be written as 
\begin{align}
    &\dot{\vb{Z}} = -\vb{v} -\gamma \vb{Z} +\lambda \nu \int \dslash{q} i \vb{q} V_{-q}\varphi_q e^{i \vb{q}\cdot \vb{Z}} + \bm{\xi}, \label{eq:frame_z} \\[2mm]
    &(\partial_t+\alpha_q - i\vb{q}\cdot \vb{v}) \varphi_q  = \lambda D q^\alpha V_q e^{-i \vb{q}\cdot \vb{Z}} + \eta_q,\label{eq:frame_varphi} 
\end{align}
with $\alpha_q$ given in \cref{eq:tau_phi}.
Note that, for $\lambda=0$, the evolution equation~\eqref{eq:frame_varphi} for $\varphi_q$ (with $\vb{v}\neq \vb{0}$) is formally the same as that for $\phi_q$ in a fixed reference frame (i.e., with $\vb{v} = \vb{0}$ --- see, c.f., \cref{eq:free_field}), up to a shift $\alpha_q\mapsto (\alpha_q - i\vb{q}\cdot \vb{v})$. 
Accordingly, its solution in the steady state is 
the same as 
the equilibrium one reported in \cref{par:freefield},
upon replacing the equilibrium correlator $C_q\z(t)$ and the free-field susceptibility $\chi_q\z(t)$ therein with 
\begin{align}
    &C_q(t) = \frac{T}{q^2+r} e^{-(\alpha_q-i\vb{q}\cdot \vb{v})\abs{t}} \equiv e^{i\vb{q}\cdot \vb{v}\abs{t}}C_q\z(t) , \label{eq:field_correlator} \\
    &\chi_q(t) = Dq^\alpha G_q(t) , \label{eq:field_susceptibility} \\
    &G_q(t) = e^{-(\alpha_q-i\vb{q}\cdot \vb{v})t}\Theta(t) = e^{i\vb{q}\cdot \vb{v}t}G_q\z(t) ,
    \label{eq:field_response}
\end{align}
where $\Theta(t)$ indicates the Heaviside theta function.
%
We will make use of these expressions in what follows.

At long times, we expect the system to reach a stationary state with $\langle\dot{\vb{Z}}\rangle_\T{ss} = 0$ and $\expval{\partial_t\varphi_q }_\T{ss} = 0$ in which, according to Eqs.~\eqref{eq:frame_z} and \eqref{eq:frame_varphi},
\begin{align}
    \expval{\vb{Z}}_\T{ss} &= -\vb{v}/\gamma +\frac{\lambda}{\kappa} \int \dslash{q} i \vb{q} V_{-q} \expval{\varphi_q e^{i \vb{q}\cdot \vb{Z}}}_\T{ss} , \label{eq:z_ss} \\
    \expval{\varphi_q}_\T{ss}  &= \frac{\lambda D q^\alpha V_q \expval{e^{-i \vb{q}\cdot \vb{Z}}}_\T{ss}}{\alpha_q - i\vb{q}\cdot \vb{v}}. \label{eq:varphi_ss} 
\end{align}
Due to the coupling between the field $\varphi_q$ and the particle coordinate $\vb{Z}$, it is difficult in general to evaluate the terms $\expval{\varphi_q e^{i \vb{q}\cdot \vb{Z}}}_\T{ss}$ and $\expval{e^{-i \vb{q}\cdot \vb{Z}}}_\T{ss}$ which appear in \cref{eq:varphi_ss,eq:z_ss}. 
For example, in \ccite{Venturelli_2022,wellGauss,Venturelli_2022_2parts} this has been achieved by a perturbative expansion in increasing powers of the weak coupling $\lambda$. 
In particular, the properties of the nonequilibrium stationary state of the system investigated here and predicted with that approach are analyzed quantitatively in Ref.~\cite{heat}, also based on numerical simulations. Here we discuss only 
some of the qualitative features \rev{that} emerge from these studies and which are relevant in the present context.

\begin{figure}[t]
    \centering 
    \includegraphics[width=\columnwidth]{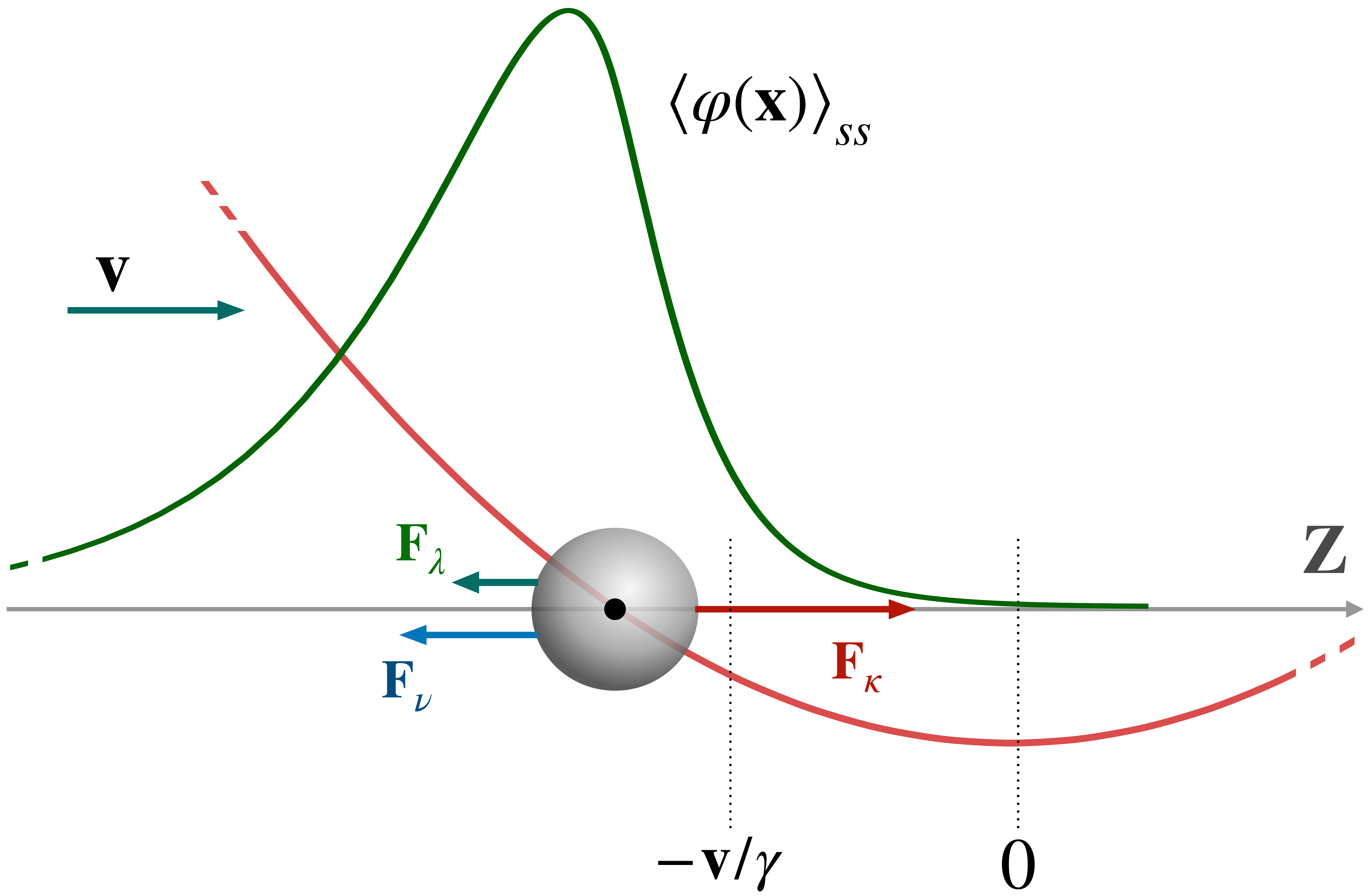}
    \caption{Sketch of the system in its nonequilibrium steady state, attained at long times in the comoving frame of reference with $\vb{Z}=\vb{Y}-\vb{v}t$. The field rearranges around the particle forming a \textit{shadow} (see \cref{eq:varphi_ss}), represented here in spatial dimensionality $d=1$ (green line). The particle is subject to the attractive force $\vb{F}_\lambda(\vb{Z})$ directed towards the shadow (and due to the field, see \cref{eq:frame_z}), to the friction force which is, on average, $\expval*{\vb{F}_\nu} = - \vb{v}/\nu$, and to the restoring force $\vb{F}_\kappa= -\kappa \vb{Z}$ due to the harmonic trap (red parabola). The steady-state position of the particle (see \cref{eq:z_ss}) results from the balance $\expval*{\vb{F}_\kappa}= \expval*{\vb{F}_\nu}+\expval*{\vb{F}_\lambda}$. For $\lambda=0$ the field and the particle are decoupled, so that $\vb{F}_\lambda=0$ and the steady-state position reduces to $\expval*{\vb{Z}}=-\vb{v}/\gamma$.
    }
    \label{fig:shadow}
\end{figure}

In the steady state, the average field profile $\expval{\varphi(\vb{x})}_\T{ss}$ (obtained from the inverse Fourier transform of $\expval{\varphi_q}_\T{ss}$) is enhanced in correspondence of the particle position, and is stretched in the direction opposite to the particle motion: 
we will refer to this field configuration as the \textit{shadow},
and we represent it schematically in \cref{fig:shadow}.
Note that, by using \cref{eq:Hint}, the term $\propto \lambda$ on the r.h.s.~of \cref{eq:EOM_frame_Z} can be written as $\lambda\nu \int {\rm d}^d\vb{z} \vb{\nabla}\varphi(\vb{z})V(\vb{z}-\vb{Z})$.
Accordingly, for $\lambda\neq 0$, the particle is subject to a force that pushes it towards the maximum of the shadow.
In the stationary state, this force adds up to the friction force in counterbalancing the restoring force exerted by the harmonic trap. 
Using perturbative arguments \cite{heat,Demery_2019}, one deduces that in general the field is responsible for the emergence of an additional (non-linear) friction acting on the dragged particle. Accordingly, the equilibrium position of the particle is further displaced to the \textit{left} with respect to the value $\expval{\vb{Z}} = -\vb{v}/\gamma$ it would have in the absence of the field (\ie, for $\lambda=0$ --- see \cref{eq:z_ss}). 
Note that the formation of the shadow is due to the response of the field to the passage of the particle, an aspect which is usually neglected in models used to describe the passive advection of a particle by a fluid flow \cite{Shraiman_2000, Falkovich_2001}.

In the present work we are primarily interested in exploring the effect of the field when the coupling $\lambda$ is relatively strong, and thus we will adopt a different approach compared to that used in \ccite{Venturelli_2022,wellGauss,Venturelli_2022_2parts,heat}. In particular, we will focus first on the \textit{noiseless} limit of the dynamics, \ie, the limit in which the amplitude $T$ of the stochastic noises $\bm{\xi}(t)$ and $\eta(\vb{x},t)$ is set to zero. This allows one to determine an analytic expression of the particle trajectory $\vb X(t)$ for generic values of $\lambda$. The effect of thermal noise when $T\neq 0$ will then be added perturbatively in \cref{sec:thermal}.

\section{Noiseless limit}
\label{sec:noiseless}

In the absence of thermal noise (\ie, for $T=0$), 
the equations of motion \eqref{eq:frame_z} and \eqref{eq:frame_varphi} become deterministic and no fluctuations occur. Accordingly, $\expval{\vb{Z}}_\T{ss} = \vb{Z}\ss$, 
$\expval{\varphi_q}_\T{ss} = \varphi_q\ss$, while
$\expval{\varphi_q e^{i \vb{q}\cdot \vb{Z}}}_\T{ss} = \varphi_q\ss e^{i \vb{q}\cdot \vb{Z}\ss}$ and 
$\expval{e^{-i \vb{q}\cdot \vb{Z}}}_\T{ss} = e^{-i \vb{q}\cdot \vb{Z}\ss}$.
Then, by using \cref{eq:z_ss,eq:varphi_ss} in the steady state,
one readily finds that 
\begin{align}
    \vb{Z}\ss &= -\vb{v}/\gamma +\frac{\lambda^2 D}{\kappa} \int \dslash{q} i \vb{q}  \frac{q^\alpha |V_q|^2}{\alpha_q - i\vb{q}\cdot \vb{v}}, \label{eq:z_ss_noiseless} \\
    \varphi_q\ss &= \frac{\lambda D q^\alpha V_q \exp[-i \vb{q}\cdot \vb{Z}\ss]}{\alpha_q - i\vb{q}\cdot \vb{v}}. \label{eq:varphi_ss_noiseless} 
\end{align}
Equation~\eqref{eq:varphi_ss_noiseless} provides the expression of the shadow in the absence of thermal noise.
As anticipated in \cref{sec:intro}, we aim to describe
the motion of the particle after it is suddenly displaced, at time $t=t_0$, from the position it assumes in the stationary state. 
In order to do this, one can solve \cref{eq:frame_varphi} (where $\eta_q=0$ in the limit we are interested in) by assuming that the field
configuration
at time $t=t_0$ 
is the one corresponding to its stationary state
--- \ie, $\varphi_q(t=t_0)=\varphi_q\ss$ is used as the initial condition of the dynamics. The resulting evolution of the field is thus given by 
\begin{equation}
    \varphi_q(t) = G_q(t-t_0)\varphi_q\ss + \lambda V_q \int_{t_0}^t \dd{s} \chi_q(t-s) e^{-i\vb{q}\cdot \vb{Z}(s)} ,
    \label{eq:varphi_risolto}
\end{equation}
where $\chi_q(t)$ and $G_q(t)$ are the field susceptibility and response propagator introduced in \cref{eq:field_response,eq:field_susceptibility}, respectively.
Equation \eqref{eq:z_ss_noiseless} suggests the natural change of reference frame, in which the origin of the coordinate system corresponds to $\vb{Z}\ss$. Accordingly, we introduce $\vb{X}\equiv \vb{Z}-\vb{Z}\ss$, so that the resting position of the particle is $\vb{X}=0$ in the stationary state (as depicted in \cref{fig:setup}). 
By substituting $\varphi_q(t)$ found in \cref{eq:varphi_risolto} into \cref{eq:frame_z} with $\bm{\xi}= \vb{0}$, we obtain the effective equation
\begin{align}
        &\dot{\vb{X}}(t) = -\vb{v} -\gamma \left[ \vb{X}(t)+ \vb{Z}\ss\right] \label{eq:effective} \\
    &\quad +\lambda^2 \nu \int \dslash{q}  \frac{i \vb{q}  |V_q|^2}{\alpha_q - i\vb{q}\cdot \vb{v}} \chi_q(t-t_0) e^{i \vb{q}\cdot \vb{X}(t)} \n \\
    &\quad +\lambda^2 \nu \int_{t_0}^t \dd{u} \int \dslash{q}  i \vb{q} |V_q|^2  \chi_q(t-u) e^{i \vb{q}\cdot \left[ \vb{X}(t)-\vb{X}(u) \right]}. \n
\end{align}
This non-linear equation with memory cannot be generically solved. 
However, further analytical progress can be made by assuming that the particle is actually perturbed by a small, sudden displacement $\vb{X}_0$ away from its resting position, as sketched in \cref{fig:setup}. 
Under this assumption, it is possible to linearize \cref{eq:effective} around $\vb{X}=0$, which leads (upon using \cref{eq:z_ss_noiseless}) to
\begin{align}
    \dot{X}_j(t) =& -X_j(t) \left[ \gamma+
    \lambda^2 \nu D \int \dslash{q}  \frac{q_j^2 q^\alpha |V_q|^2}{\alpha_q - i\vb{q}\cdot \vb{v}} \right]     \label{eq:linearized} \\
    &\, +\lambda^2 \nu \int_{t_0}^t \dd{u} \int \dslash{q} q_j^2 |V_q|^2  \chi_q(t-u) X_j(u), \n
\end{align}
for $j=1,\dots,d$.
Let us now introduce the memory kernel
\begin{equation}
    \Gamma_j(t) \equiv \lambda^2 \nu \int \dslash{q} q_j^2 |V_q|^2  \chi_q(t)
    \label{eq:memory_kernel}
\end{equation}
and its Laplace transform
$
    \hat \Gamma_j (s) = \int_0^\infty \dd{t} e^{-st} \Gamma_j(t)
$; in terms of these quantities, the linearized equation of motion \eqref{eq:linearized} can be written in the compact form
\begin{equation}
    \dot{X}_j(t) = -X_j(t) \left[ \gamma+ \hat \Gamma_j(0) \right] + \int_{t_0}^t \dd{u} \Gamma_j(t-u) X_j(u). \label{eq:linearized_compact}
\end{equation}
We recognize \cref{eq:linearized_compact} as the noiseless limit of an overdamped generalized Langevin equation \cite{Mori_1965,Zwanzig_book}.
By setting $t_0=0$, the solution of the latter equation with initial condition $X_j(t=0)=X_0$ can be conveniently expressed in Laplace space as
\begin{equation}
    \hat X_j(s) = \frac{X_0}{s+\gamma -\left[ \hat \Gamma_j(s)-\hat \Gamma_j(0) \right]},
    \label{eq:X(s)}
\end{equation}
where, as in the case of $\hat \Gamma_j$ after \cref{eq:memory_kernel}, $\hat X_j(s)$ stands for the Laplace transform of $X_j(t)$.

\subsection{The memory kernel $\Gamma(t)$}

The dynamics of $X_j(t)$ is determined by the analytic structure of the function $\hat \Gamma_j(s)$ in the complex plane, which we discuss here. 
For later convenience, it is useful to introduce the following timescales:
\begin{align}
    \tau_R &\equiv R^{z}/D, \label{eq:tau_R} \\
    \tau_v &\equiv R/v. \label{eq:tau_v}
\end{align} 
The first timescale $\tau_R$ is the time taken by a critical field to relax over a distance of order $R$: this can be seen by using \cref{eq:tau_phi} with $r=0$, $q\simeq 1/R$, and $z\equiv 2+\alpha$. (We recall that $R$ enters as a length scale in $V(\vb{x})$, and plays the role of the radius of the particle described by $V(\vb{x})$.) The second timescale $\tau_v$ represents, instead, the time taken by the moving trap to cover a distance of order $R$; equivalently, $\tau_v^{-1}$ estimates the shear rate near the driven particle \cite{Berner2018}.

By rescaling momenta as $p=qR$ in \cref{eq:memory_kernel} and evaluating the Laplace transform, we easily obtain
\begin{equation}
    \hat \Gamma_j(s) = \frac{\lambda^2 \nu}{R^d} f(s;\tau_R,\tau_v,R,\xi) = \frac{\lambda^2 \nu}{R^d} \tilde f(s\tau_R,\tau_R/\tau_v,R/\xi) ,
    \label{eq:memory_kernel_scaling}
\end{equation}
where the prefactor $\lambda^2 \nu /R^d$ has the physical dimensions of an inverse time, while $\tilde f$ is a dimensionless scaling function defined as
\begin{align}
    \tilde f(\theta_1,\theta_2, \theta_3) = \int \dslash{y}  \frac{y_j^2 y^\alpha |V_{y/R}|^2}{\theta_1+y^\alpha\left(y^2+\theta_3^2\right) - i \theta_2 \vb{y}\cdot \vu{v}} . \label{eq:scaling_memory}
\end{align}
Note that $V_{y/R}$ is in fact $R$-independent by construction (see, e.g., \cref{eq:potential_peak}).
Moreover, the timescale $\tau_\kappa$ (see \cref{eq:tau_kappa}) which determines the relaxation time of the  particle (decoupled from the field) in the harmonic trap does not enter the memory kernel, which thus describes solely the interaction between the medium and the particle.
By substituting \cref{eq:memory_kernel_scaling} into \cref{eq:X(s)}, one eventually finds
\begin{equation}
    \hat X_j(s) = \frac{X_0/\gamma}{1+s/\gamma -g\left[ f(s)-f(0) \right]},
    \label{eq:X(s)_rephrased}
\end{equation}
where we introduced the dimensionless coupling constant
\begin{equation}
    g \equiv \frac{\lambda^2}{\kappa R^d},
    \label{eq:dimensionless_coupling}
\end{equation}
and where we simplified the notation by explicitly indicating only the dependence on $s$ of $f$ introduced in \cref{eq:memory_kernel_scaling}. 
The coupling constant $g$ can be used to quantify the effect of the interaction with the medium on the particle dynamics.
We note that $\hat X(s)$ in \cref{eq:X(s)_rephrased} satisfies the initial value theorem for Laplace transforms \cite{Schiff_1999}, \ie,
\begin{equation}
    \lim_{s\to \infty} s \hat X(s) = X(t=0^+) = X_0 ,
    \label{eq:IVT}
\end{equation}
as expected --- indeed, one can check that $f(s)\sim 1/s$ for large $s$.

In order to 
get physical insight into the dynamics of the particle, it is convenient to consider the case in which the timescales $\tau_R$ and 
$\tau_\kappa$, which $\hat X_j$ depends on via $f$, are well separated. This is actually achieved in the \textit{strong-confinement limit} \cite{Demery_2019}, defined as the limit in which $\tau_\kappa$, determined by the harmonic trap, is shorter than the typical relaxation time $\tau_R$ of the field, i.e., $\tau_R \gg \tau_\kappa$ or, equivalently,
\begin{equation}
    \rho \equiv \tau_R/\tau_\kappa = \gamma \tau_R  \gg 1
    \label{eq:rho}
\end{equation}
(see \cref{eq:tau_kappa}).
In this limit we will focus on the dynamics occurring at times $t\gg \tau_\kappa$, so that $\tau_\kappa$ is indeed the smallest timescale in the problem. A convenient way of singling out the behavior in this temporal regime 
is to consider, in \cref{eq:X(s)}, the formal limit $\gamma\to\infty$ and thus
\begin{equation}
    \hat X\sc(s) \equiv \lim_{\gamma \to \infty} \gamma \hat X(s) = \frac{X_0}{1 -g\left[ f(s)-f(0) \right]},
    \label{eq:strong_confinement}
\end{equation}
the analysis of which is simplified by the fact that $\hat X^{(\T{sc})}(s)$ depends on $s$ only via the function $f(s)$. 
As a drawback of this approach, $\hat X^{(\T{sc})}(s)$ defined above no longer satisfies the initial value theorem and, as a consequence, its inverse Laplace transform $X^{(\T{sc})}(t)$ diverges in the initial temporal region $t\leq \tau_\kappa$. Beyond this initial regime, however, the functions $X^{(\T{sc})}(t)$ and $\gamma X(t)$ are expected to agree quantitatively (as discussed in, c.f., 
\rev{\cref{sec:discussion}}).
%

\begin{figure}[t]
    \centering \includegraphics[width=\columnwidth]{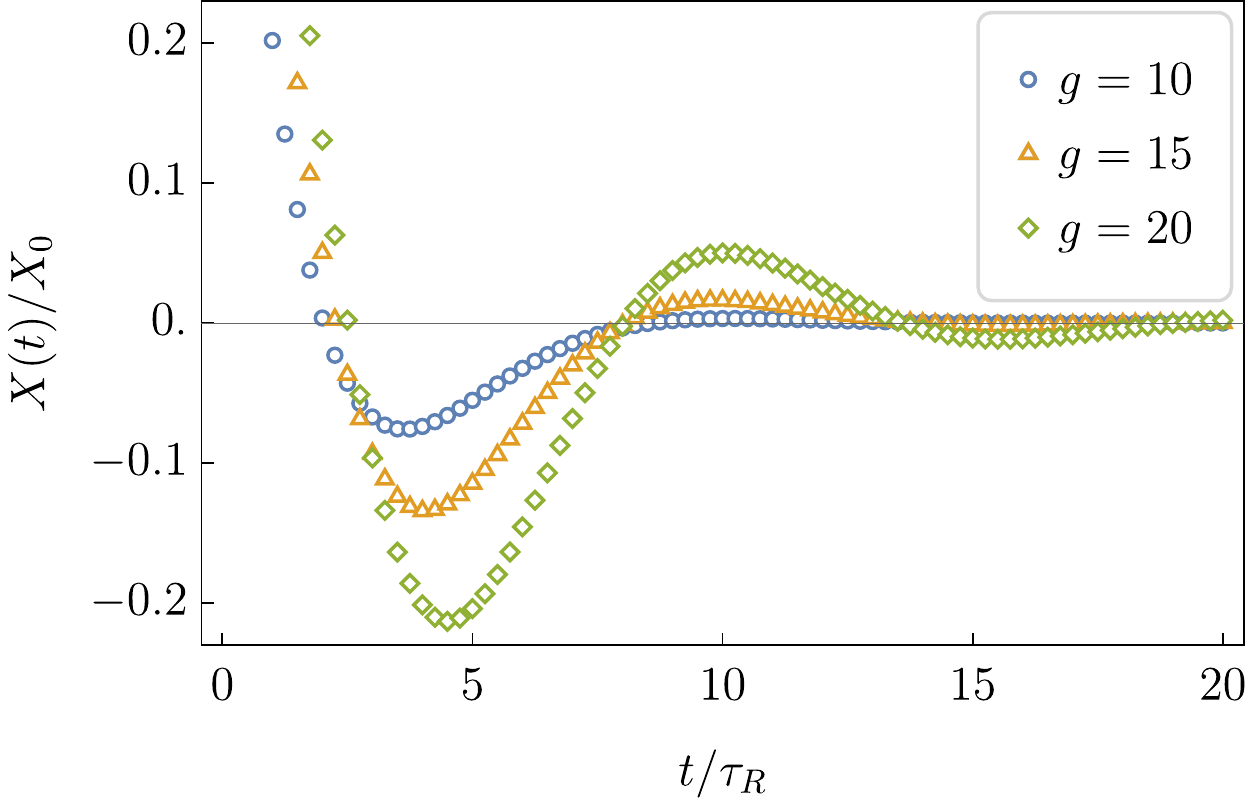}
    \caption{Evolution of the particle position $X(t)$ as a function of time $t$, in the noiseless limit. At $t=0$ the particle is released from the position $X(t=0^+) = X_0$ away from the 
    steady-state position $X=0$.
    The displayed symbols are obtained from the numerical Laplace inversion of $\hat X\sc(s)$ in \cref{eq:strong_confinement}, corresponding to the strong-confinement limit. The plot refers to model A at criticality in spatial dimensionality $d=1$ (see \cref{sec:modelA}), and was obtained by fixing the Weissenberg number $w=0.75$ (see \cref{eq:weissenberg}), while varying the coupling strength $g$ (see \cref{eq:dimensionless_coupling}), with $X_0=1$. The so-obtained $X\sc (t)$ differs from the actual $\gamma X(t)$ only at short times $t<\tau_\kappa$, where the former diverges (while the latter tends to $\gamma X_0$ --- see \cref{eq:IVT}).
    }
    \label{fig:sample-oscillations}
\end{figure}

\subsection{The case of model A}
\label{sec:modelA}

To make further progress with our analysis, we focus here on the one-dimensional case $d=1$, with the field poised at its critical point $r=0$ (further below we consider also the case $r>0$).
In addition, we choose an exponential interaction potential as in \cref{eq:potential_peak_fourier}, 
which takes a particularly simple form in Fourier space, namely
\begin{equation}
    V_q = (1+q^2R^2)^{-1}.
\label{eq:potential_peak_fourier}
\end{equation}
This choice renders the expressions below more amenable to analytical manipulation.
In fact, the resulting memory kernel in \cref{eq:memory_kernel_scaling} becomes 
\begin{equation}
    \hat \Gamma(s) = \frac{\lambda^2 \nu}{R} \int_\mathbb{R} \frac{\dd{q}}{2\pi} \frac{q^z}{(1+q^2)^2(q^z-iq \tau_R / \tau_v+s\tau_R)},
    \label{eq:Gamma-crit-1d-exp}
\end{equation}
where we dropped the subscript $j$ from $\hat \Gamma_j(s)$ since we are considering $d=1$.
In the Gaussian model A, the dynamical exponent $z$ equals 2, so that the integrand in $\hat \Gamma(s)$ presents two simple poles in $q=\pm i$ and two additional poles in
\begin{equation}
    q_\pm = i \left[ \frac{\tau_R}{2\tau_v} \pm \sqrt{ s\, \tau_R + \left( \frac{\tau_R}{2\tau_v} \right)^2 }\; \right] \equiv i\, [w\pm \beta(s)].
    \label{eq:poles}
\end{equation}
For later convenience, we parameterized these latter two poles as indicated above, with
\begin{equation}
    w \equiv \frac{\tau_R}{2\tau_v} = \frac{v R^{z-1}}{2 D}
    \label{eq:weissenberg}
\end{equation}
(see Eqs.~\eqref{eq:tau_R} and \eqref{eq:tau_v}), and $\beta(s)\equiv \sqrt{s\, \tau_R+w^2}$.
In the context of microrheology experiments conducted in viscoelastic media,
one usually identifies the Weissenberg number $\T{Wi}\equiv \tau_s/(2\tau_v)$, where $\tau_s$ is the typical relaxation timescale of the medium.
For a critical field this timescale is actually provided by $\tau_R$ (see \cref{eq:tau_R}), and therefore the parameter $w$ introduced in \cref{eq:weissenberg} above is readily identified with the Weissenberg number $\T{Wi}$ of the system under investigation here.
By using complex integration, one then finds that $\hat \Gamma(s)$ in \cref{eq:Gamma-crit-1d-exp} can be expressed as 
\begin{equation}
    \hat \Gamma(s) = \frac{\lambda^2 \nu}{R}\,\frac{\beta(s)[1+\beta(s)]^2-w^2[2+\beta(s)]}{4\beta(s)[1+\beta(s)+w]^2[1+\beta(s)-w]^2}.
    \label{eq:modelA_1d_critical}
\end{equation}
%
%
%
%
%
This expression implies $\hat \Gamma(s=0)= \lambda^2 \nu/[ 4R(1+2w)^2 ]$, which can be inserted into \cref{eq:X(s)} together with $\hat \Gamma(s)$ given above in  order to obtain an analytical expression for $\hat X(s)$. The latter can then be inverted numerically to determine $X(t)$. 
An example of the resulting $X(t)$ is shown in \cref{fig:sample-oscillations}, which refers to the strong-confinement limit, while a comparison with numerical simulations is presented in, c.f., \cref{sec:simulation}. The oscillatory character of this $X(t)$ is clearly visible from the figure and it can be amplified by increasing the coupling strength $g$ (a systematic analysis of this dependence is presented in the next subsection).

By inspecting \cref{eq:scaling_memory,eq:memory_kernel_scaling}, we finally note that the expression of $\hat \Gamma(s)$ for model A away from criticality (\ie, with $r>0$) can be obtained from \cref{eq:modelA_1d_critical} by means of the substitution
\begin{equation}
s \mapsto s+ \tau_\xi^{-1}, \label{eq:analytic_continuation}    
\end{equation}
where 
\begin{equation}
\tau_\xi \equiv 1/(Dr) = \xi^2/D \label{eq:tau_xi}
\end{equation} 
quantifies the relaxation timescale of the field $\varphi$ over its correlation length $\xi=r^{-1/2}$ (see \cref{eq:tau_phi} with $q\simeq \xi^{-1}$ and $z=2$).


\begin{figure*}
\centering
\subfloat[]{
  \centering
  \includegraphics[width=0.61\columnwidth]{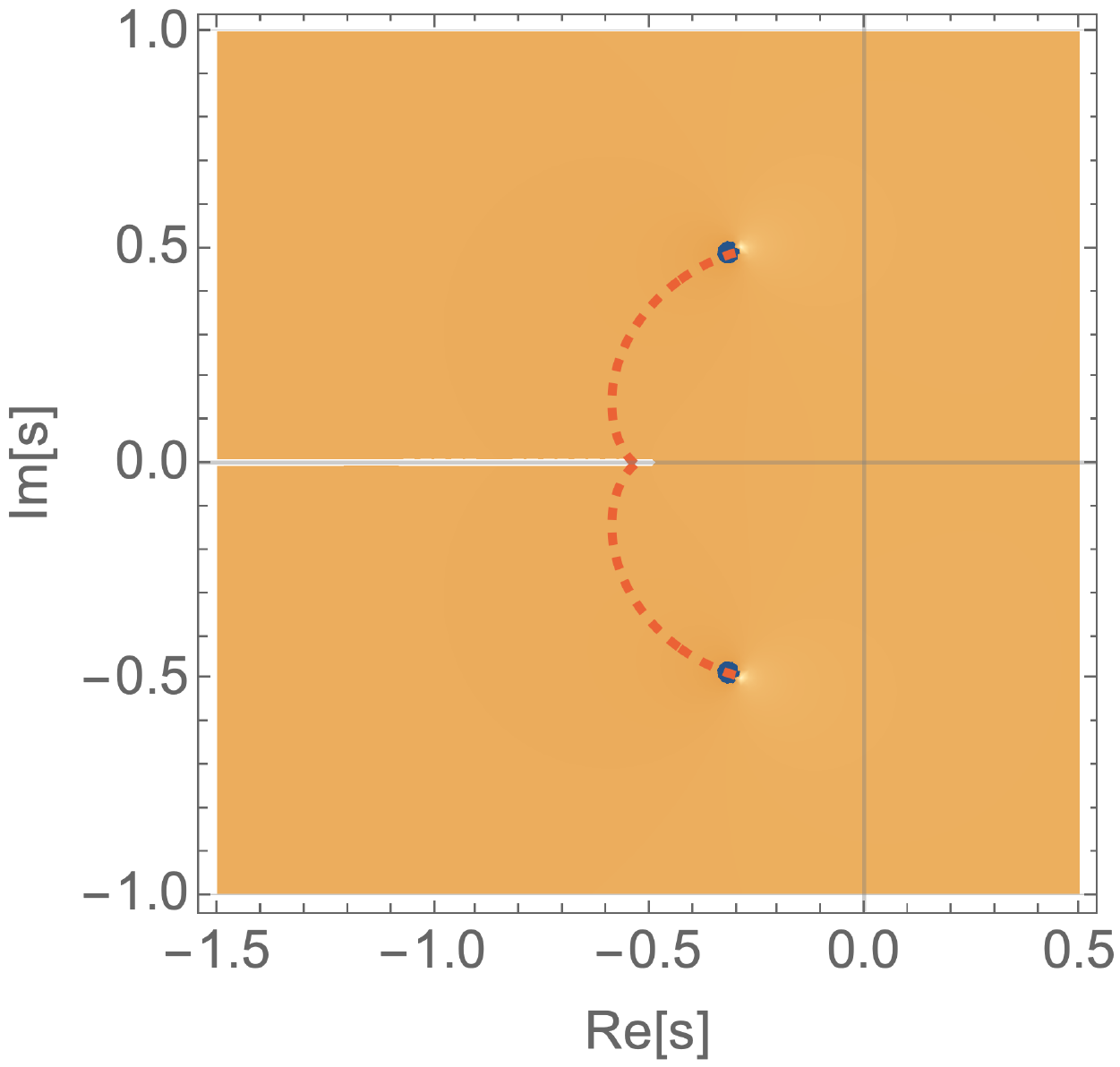}
  \label{fig:poles}
  }
  \subfloat[]{
  \centering
  \includegraphics[width=0.69\columnwidth]{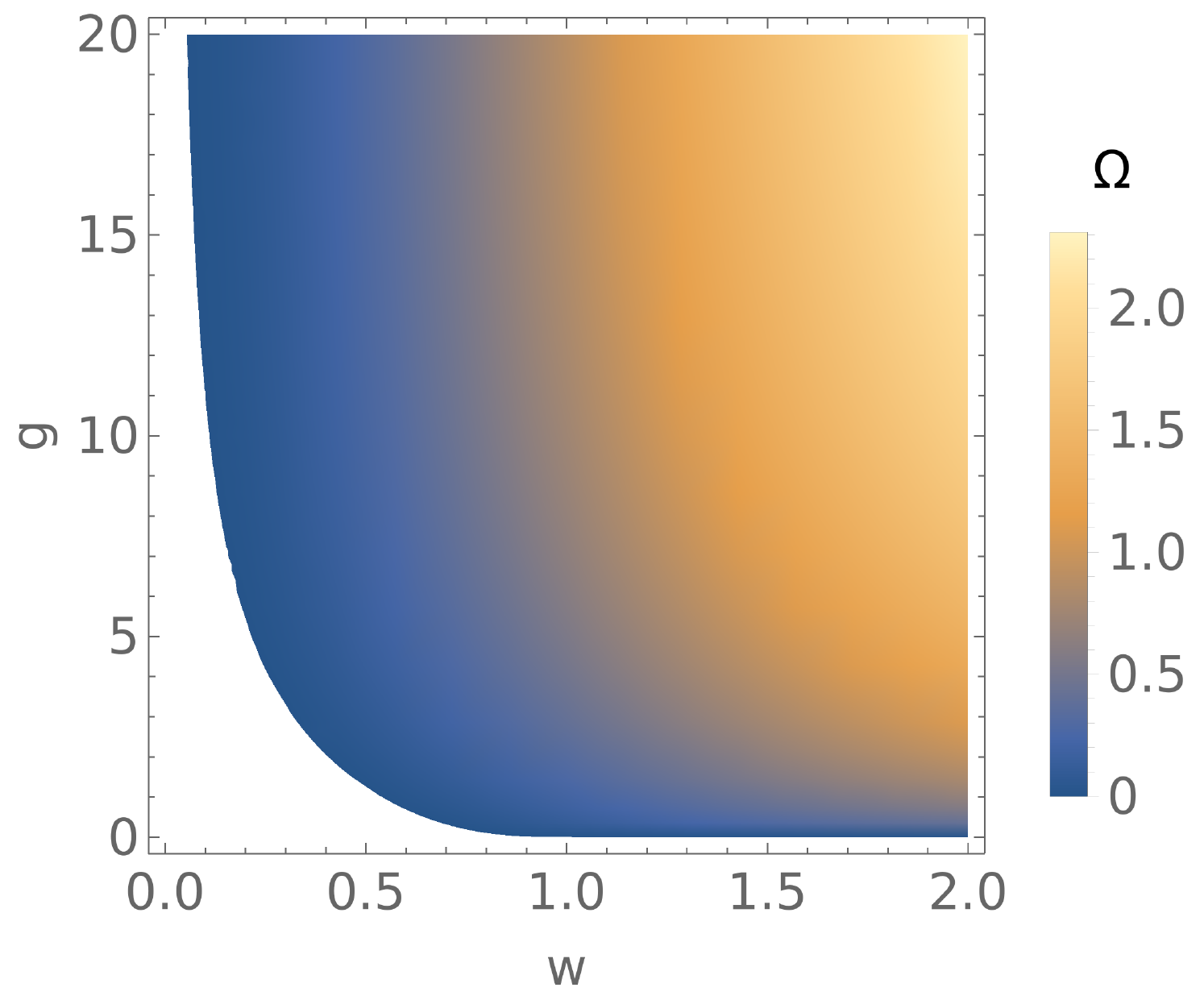}
  \label{fig:phase-diagram}
  }
  \subfloat[]{
  \centering
  \includegraphics[width=0.69\columnwidth]{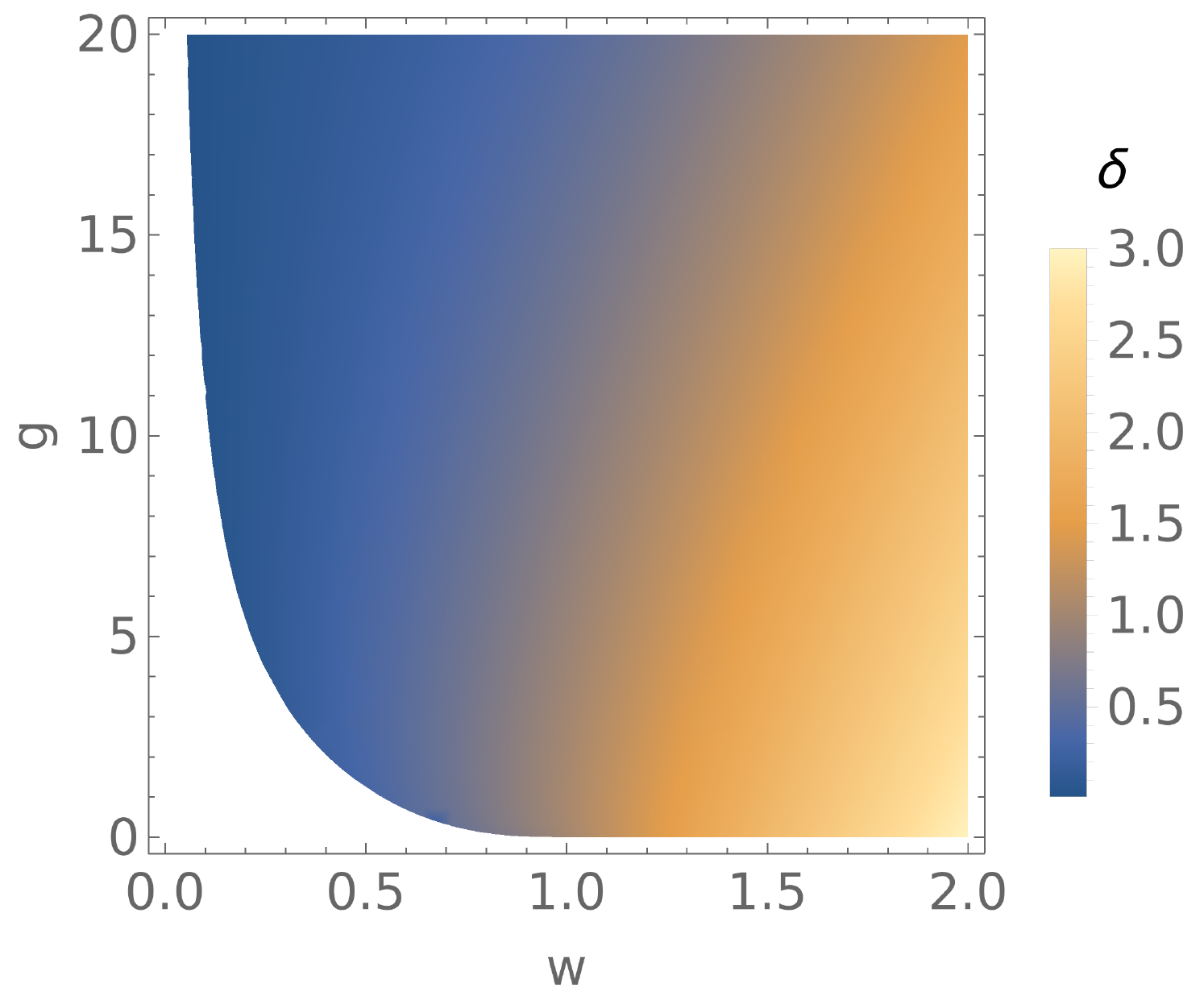}
  \label{fig:phase-diagram_2}
  }
\caption{Analytic structure of the Laplace transform $\hat X(s)$ of $X(t)$ in the complex plane $s\in{\mathbb C}$, for model A in spatial dimension $d=1$, at criticality $r=0$, and in the strong-confinement limit (see \cref{sec:phase-diagram}). 
\textbf{(a)} Plot of $\Re \{\hat X(s)\}$, 
which shows the presence of a branch cut for $\Re{s}<-w$, and a pair of complex conjugate poles $s_\pm = -\delta \pm i \Omega$ with nonzero imaginary part $\Omega>0$. The latter determines the oscillatory 
frequency $\Omega$ of $X(t)$ after a short initial transient (see also \cref{fig:sample-oscillations}). The plot corresponds to $w=0.7$, and the red dashed lines indicate the trajectories of the poles in the complex plane, which emerge out of the branch cut and move away from it upon increasing the values of $g\in[0.32,17]$. 
\textbf{(b)} Imaginary part $\Omega$ of the upper pole $s_+$ as a function of $w$ and $g$. Oscillations develop for any value of $g$ as soon as $w \gtrsim 1$, while they are absent within the white region, where there are no poles and the decay is controlled by the branch cut.
The frequency $\Omega$ is measured in units of $\tau_R^{-1}$ --- see \cref{eq:tau_R}.
\textbf{(c)} Real part $\delta$ of the poles (which controls the rate of the exponential damping of the oscillation amplitude) as a function of $w$ and $g$. Oscillations turn out to be increasingly damped upon increasing the value of $w$.
As in panel (b), no poles emerge within the white region in this plane.
}
\label{fig:analytic-structure}
\end{figure*}

\subsection{Relaxation in the strong-confinement limit}
\label{sec:phase-diagram}

In the strong-confinement limit introduced in \cref{eq:strong_confinement}, the analytic properties of $\hat X\sc(s)$ are completely determined by those of the memory kernel $\hat \Gamma(s)$ and, in particular, of the associated function $f(s)$ introduced in \cref{eq:memory_kernel_scaling}. This kernel was
specialized
in \cref{eq:modelA_1d_critical} 
to the case of model A at criticality ($r=0$), while for $r>0$ one can use the change of variable indicated in \cref{eq:analytic_continuation}. The latter implies 
$\hat X_r\sc(s) = \hat X_{r=0}\sc(s+Dr)$, and therefore
\begin{equation}
    X_r\sc(t) = \int_{c-i\infty}^{c+i\infty} \frac{\dd{s}}{2\pi i} e^{s t}  \hat X_r\sc(s) \label{eq:bromwich}
    = e^{-D r t} X_{r=0}\sc(t) .
\end{equation}
Above we highlighted the dependence of $X\sc(t)$ on $r$ via a subscript. The inverse Laplace transform of $\hat{X}\sc(s)$ in the previous expression is obtained, as usual, by performing the Bromwich integral along a vertical line \rev{that} is on the left of the leftmost pole of the integrand in the complex plane.
Accordingly, in model A, the dynamical properties of $X\sc(t)$ in the off-critical case $r\neq 0$ are the same as in the critical case $r=0$, up to an additional exponential decay factor $\exp(- t/\tau_\xi)$ (see \cref{eq:tau_xi}).

A second remarkable feature of $\hat X\sc (s)$ 
is that it depends on $s$ only via the combination $s \, \tau_R$ --- see \cref{eq:memory_kernel_scaling,eq:scaling_memory}, and the definition of $\tau_R$ in \cref{eq:tau_R}. 
Taking the inverse Laplace transform of $\hat X\sc (s)$ as in \cref{eq:bromwich} and changing the integration variable as $s'\equiv s \tau_R$, it follows that
\begin{equation}
    X\sc (t) = \tau_R^{-1} X\sc (t/\tau_R).
\end{equation}
We deduce that rescaling $s\, \tau_R \mapsto s$ in $\hat X\sc(s)$ simply corresponds to measuring time $t$ in units of $\tau_R$. 
Furthermore, the explicit dependence on $R$ of $\hat \Gamma (s)$, which occurs in \cref{eq:memory_kernel_scaling} only via $R/\xi$, is lost at criticality ($\xi\to\infty$).
Accordingly, the resulting $\hat X\sc(s)$ eventually depends only on the pair of parameters $(w,g)$ --- see \cref{eq:poles,eq:modelA_1d_critical,eq:weissenberg,eq:dimensionless_coupling}.

Let us then focus on the analytic structure of $\hat X\sc (s)$ in the complex plane $s\in\mathbb{C}$, 
for $r=0$ and with $\tau_R \equiv 1$. 
First, from \cref{eq:poles} we infer the presence of a branch cut along the real axis for $\Re{s}< -w$, as shown in \cref{fig:poles}. 
The exact position of the poles of 
$\hat X\sc(s)$ 
in \cref{eq:strong_confinement}
cannot be determined analytically; however, they are easily found numerically.
Indeed, the plot of $\T{Re} \{\hat X(s)\}$ in the complex plane, shown in \cref{fig:poles} as a colormap,
reveals the presence of a pair of complex conjugate poles in $s_\pm=-\delta \pm i \Omega$, with $\Omega\ge 0$. 
The red dashed line in the plot indicates the trajectory of these poles upon varying $g$ at fixed $w<1$. The poles appear for a small $g =g^*> 0$, in the vicinity of the origin $s=-w$ of the branch cut, and have a vanishing imaginary part $\Omega= 0$; 
upon increasing $g$, 
they depart from the branch cut and acquire a nonzero imaginary part $\Omega>0$. As $g$ is further increased, the two poles eventually move to the right of the branching point (i.e., $|\delta| <|w|$), and thus they become the \textit{dominant} singularities.
The presence of a dominant complex pole in the analytic structure of $\hat X\sc (s)$ implies the emergence of an oscillatory behavior of $X\sc(t)$ at long times, with frequency $\Omega$ (see \cref{app:tauberian} for additional details).
These are the oscillations featured in \cref{fig:sample-oscillations}, where we plotted $X\sc(t)$ (obtained via numerical inversion of the analytical solution for $\hat X\sc(s)$) for increasing values of the coupling strength $g$, while keeping $w$ fixed.

Figure~\ref{fig:phase-diagram} shows the oscillation frequency $\Omega$ as a function of the values of the parameters $(w,g)$. Within the white region of the plot there are no poles, and thus no oscillations occur.
Even for small values of $g$, instead, complex poles appear and oscillations are seen to develop as soon as $w\gtrsim 1$; moreover, $\Omega$ is in general an increasing function of $w$, for any fixed value of $g$.
We recall that $w=\T{Wi}$ measures the ratio between the relaxation time of the medium and the timescale $\tau_v$ set by the moving trap (see \cref{eq:weissenberg}). This suggests a way to rationalize 
the ``dynamical phase diagram'' in \cref{fig:phase-diagram}. 
Indeed, for small $g$ and sufficiently large values of the dragging speed $v\propto w$ (see \cref{eq:tau_v,eq:weissenberg}), the field is no longer able to quickly rearrange around the instantaneous position assumed by the particle at a given time: the non-Markovian interplay between the dynamics of the particle and the field shadow is then at the origin of the complex oscillatory behavior of $X(t)$. The effects of this interplay become increasingly prominent upon increasing the coupling strength $g$, so that at large $g$ one observes an oscillatory behavior even for $w\lesssim 1$. 

Conversely, these oscillations are increasingly damped upon increasing the dragging speed, i.e., for $w\gg 1$. This is shown in \cref{fig:phase-diagram_2}, where we plot the real part $\delta$ of the dominant pole as a function of $(w,g)$ --- indeed, $\delta$ controls the long-time exponential decay of $X(t)$ (see \cref{app:tauberian}). 
To understand the damping at large $w\propto v$, we first note that
the shape
of the shadow $\varphi\ss(\vb{x})$ is given by the Fourier transform of $\varphi_q\ss$ in \cref{eq:varphi_ss}: upon inspection, the latter shows that the amplitude of the shadow itself decreases upon increasing $v$ \cite{heat}. 
This is expected, since the finite relaxation time of the field $\varphi$
does not allow $\varphi$ to react instantaneously to the passage of the particle, and thus at a very large speed $v$ the shadow cannot build up at all. The damping of the oscillations at large values of $w$ thus simply reflects these facts.

We emphasize that no poles emerge in $\hat X(s)$ within the white region in the $(w,g)$-plane in \cref{fig:phase-diagram,fig:phase-diagram_2}. Correspondingly, the long-time behavior of $X(t)$ in that region is determined solely by the branch cut (see \cref{fig:poles}): as we recall in \cref{app:tauberian}, this generically implies that $X(t)$ decays monotonically as $X(t)\sim t^{-a} \exp(-b t)$, for some positive constants $a$ and $b$ (see \cref{eq:branch_point}).
Conversely,
upon increasing $g$ far beyond the values \rev{that} \cref{fig:poles} refers to, the real part  $-\delta$ of the poles $s_\pm$ eventually becomes positive. This would imply an unbounded (oscillatory) growth of $X(t)$ at long times (see \cref{app:tauberian} for details), and thus it signals the breakdown of the linear-response approximation within which such solution has been derived.

\subsection{\rev{How generic are these oscillations?}}
\label{sec:discussion}

Beyond the strong-confinement limit discussed  \rev{in Sec.~\ref{sec:phase-diagram} above,}
\ie, upon decreasing the value of the parameter $\gamma$, new poles eventually appear in the complex-$s$ plane shown in \cref{fig:poles}.
Although the precise value of $\Omega$ at a certain point $(w,g)$ of the plane is in general modified compared to the value it has in the strong-confinement limit $\rho \gg 1$ (see \cref{eq:rho}), we find that the oscillatory nature of the solution $X(t)$ persists, within the same range of values as in \cref{fig:phase-diagram}, down to $\rho \gtrsim 1$.
Note that, after rescaling $s'\equiv s \, \tau_R$ in \cref{eq:X(s)_rephrased}, the latter reads
\begin{equation}
    \hat X_j(s'/\tau_R) = \frac{X_0/\gamma}{1+s'/\rho -g\left[ f(s'/\tau_R)-f(0) \right]} ,
    \label{eq:X(s)_rho}
\end{equation}
showing (as expected)
that the strong-confinement limit becomes increasingly accurate as $\rho  \gg 1$ --- compare with \cref{eq:strong_confinement}. 
Moreover, 
\cref{fig:sample-oscillations,fig:analytic-structure} (together with the numerical simulations presented in, c.f., \cref{sec:simulation}) show that $X(t)$ typically decays to zero on a scale of a few tens of $\tau_R$. 
As a result, even for $\rho \lesssim 1$, the strong-confinement limit well approximates the behavior of $X(t)$ at times $t>\tau_\kappa=\gamma^{-1}$. 
Indeed, by taking 
the inverse Laplace transform of $\hat X_j(s)$ in \cref{eq:X(s)_rephrased} and calling $z\equiv st$, one obtains
\begin{equation}
    X_j(t) 
    = (\gamma t)^{-1} \int_B \frac{\dd{z}}{2\pi i} \frac{e^{z} X_0}{1+z/(\gamma t) -g\left[ f(z/t)-f(0) \right]} ,
\end{equation}
where the integration is intended along the Bromwich contour as in \cref{eq:bromwich}. The term $z/(\gamma t)$ at the denominator can be safely neglected as soon as $\gamma t\gg 1$, yielding in fact $ X_j(t) \simeq  X_j\sc(t) /\gamma$ (see \cref{eq:strong_confinement}).

Away from the critical point (i.e., for $r>0$), the damped  oscillations of $X(t)$ persist, but they are additionally suppressed by the exponential factor $\exp(-Dr t)=\exp(-t/\tau_\xi)$ (see \cref{eq:tau_xi,eq:bromwich}). 
Taking into account all the trends highlighted above, we expect that the oscillatory behavior of $X(t)$ is maximally amplified within the timescale window $\tau_\kappa< \tau_R < \tau_\xi$, where the second inequality corresponds to requiring $\xi>R$ (see \cref{eq:tau_R,eq:tau_xi}). 

Note that increasing the trap strength $\kappa$ has the effect of both increasing $\gamma$ (thus pushing the system further into the strong-confinement regime), and decreasing the effective coupling $g$ and therefore \rev{reducing} the amplitude of the oscillations (see \cref{fig:phase-diagram,fig:phase-diagram_2}). 
Accordingly, oscillations generically develop at intermediate values of $\kappa$, while they vanish both at very large and very small values of $\kappa$. This was also the case in experiments performed on colloidal particles dragged in viscoelastic media (see \ccite{Berner2018} and Fig.~5 therein).

Figure~\ref{fig:analytic-structure} additionally 
confirms that no oscillations occur if the trap is not dragged, \ie, for $v=0$ (hence $w=0$). This was also the case in the experiments of \ccite{Berner2018} involving a viscoelastic medium (see Fig.~3 therein). 
This fact also 
agrees with the analytical and numerical results of \ccite{Venturelli_2022}, 
where
the relaxation towards equilibrium of a trapped particle in contact with a near-critical Gaussian field was 
investigated
perturbatively in the coupling $\lambda$. In particular, it was found that 
$\expval{X(t)}$ decreases algebraically upon increasing time $t$ for model A at criticality, and generically for model B.
For completeness, in \cref{app:relaxation} we reconsider this problem within the noiseless but non-perturbative approach presented in this Section, and we re-derive the exponents of the long-time algebraic decay of $\expval{X(t)}$ originally reported in \ccite{Venturelli_2022}.

\rev{Finally, the emergence of oscillations in viscoelastic media reported in \ccite{Berner2018} was rationalized therein in terms of the stochastic dynamics of an underdamped harmonic oscillator. In fact, it was shown that such a simplified model (with a positive, memory-induced mass term) is able to reproduce quantitatively the oscillations} 
\rev{displayed
at long times 
by
the dragged colloidal particle.}
%
While such an 
\rev{effective model}
turns out to be inappropriate in our case due to the non-analytic behavior of the memory kernel,
in \cref{app:berner} we discuss in detail 
\rev{the comparison between our model and the}
\rev{theoretical description of} viscoelastic fluids used in \ccite{Berner2018}. 
In particular, \rev{it turns out that} the memory kernels $\Gamma(t)$ emerging in the present case and in viscoelastic media appear to be both negative at long times $t$, confirming that the \textit{negative response} of the surrounding medium (whose origin in our model has been clarified in the previous Sections) is 
\rev{actually essential}
for the emergence of the oscillating modes exhibited by the overdamped particle.

\section{Effects of thermal fluctuations}
\label{sec:thermal}

Thermal fluctuations act on the field and the particle, via the noise terms $\bm{\xi}(t)$ and $\eta_q(t)$ in \cref{eq:frame_z,eq:frame_varphi}, whenever $T\neq 0$. The presence of thermal noise represents an obstacle to the analytical derivation of the 
time-dependent relaxation 
of the particle, because it modifies the steady-state average of both the particle position $\expval{\vb{Z}}_\T{ss}$ in \cref{eq:z_ss}, and the field profile $\expval{\varphi_q}_\T{ss}$ in \cref{eq:varphi_ss}. 
Once incorporated into the effective equation of motion of the particle, the field-induced fluctuations turn out to be non-Gaussian, as we will verify shortly; in order to account for them, we shall resort below to a perturbative expansion in the coupling constant $\lambda$ (as previously done in related investigations of this model \cite{Venturelli_2022,wellGauss,Venturelli_2022_2parts,heat,Gross_2021}). 
We emphasize that the (noiseless) effective equation \eqref{eq:effective} is actually \textit{non-perturbative} in $\lambda$, and so is its solution in \cref{eq:X(s)_rephrased}. 
Expanding the dynamics for small $\lambda$ is just a computational tool to take fluctuations into account analytically, but the qualitative conclusions we reach are valid beyond the perturbative regime, as we confirm in \cref{sec:simulation} by using numerical simulations.

\subsection{Weak-coupling approximation}
\label{sec:thermal_derivation}

The effective equation \eqref{eq:effective} in \cref{sec:noiseless} was determined first by choosing the shadow state in \cref{eq:varphi_ss_noiseless} as the initial condition for the field $\varphi$ at time $t=t_0$, and then by moving to a reference frame in which the resting position of the particle corresponds to $\vb{X}=0$. 
In this Section we adopt a different strategy: instead of explicitly determining the stationary 
shadow configuration (which is difficult in the presence of thermal fluctuations), we first solve for $\varphi_q(t)$ as we did in \cref{eq:varphi_risolto}, but we impose the flat initial condition $\varphi_q(t=t_0)=0$ at the initial time $t_0$, and we take into account the contributions due to the noise.  
Plugging the result into \cref{eq:frame_z} then yields
\begin{align}
    \dot{\vb{Z}}(t) =& -\vb{v} -\gamma \vb{Z} + \bm{\xi}(t) +\nu \lambda \int \dslash{q} i \vb{q} V_{-q} e^{i \vb{q}\cdot \vb{Z}(t)} \n \\
    &\times \left[\zeta_q(t)+ \lambda V_q \int_{t_0}^t \dd{s}   \chi_q(t-s) e^{-i \vb{q}\cdot \vb{Z}(s)}  \right] ,
    \label{eq:effective_equation}
\end{align}
where the field susceptibility $\chi_q(u)$ was defined in \cref{eq:field_susceptibility}, and where we introduced the Gaussian colored noise 
\begin{equation}
    \zeta_q(t) \equiv \int_{t_0}^t \dd{s} G_q(t-s) \eta_q(s),
    \label{eq:colored_noise}
\end{equation}
which has zero mean and correlator $C_q(t)$ (see \cref{eq:field_correlator}).

Although we did not specify the stationary shadow configuration 
of the field (see \cref{fig:shadow}) as the initial condition of 
its evolution, one can convince oneself that such a configuration is inevitably recovered by taking the limit $t_0\to -\infty$, since it coincides with the nonequilibrium steady state of the system. 
The leading correction to the average particle position $\expval{\vb{Z}}$, due to thermal fluctuations, can then be 
\rev{calculated from Eq.~\eqref{eq:effective_equation} by following the steps detailed in \cref{app:correction}. This eventually yields}
\begin{align}
    &\expval{\vb{Z}}_\T{ss} = -\vb{v}/\gamma + \frac{\lambda^2}{\kappa} \int \dslash{q} i \vb{q} |V_q|^2  \label{eq:Z_ss_corr} \\
    &\times   \int_{0}^\infty \dd{u} \left[ \chi_q(u)+\nu q^2 e^{-\gamma u} C_q(u)  \right]e^{-q^2 \sigma^2(u)}  +\order{\lambda^4}, \n
\end{align}
with
\begin{equation}
    \sigma^2(u) \equiv \frac{T}{\kappa}\left( 1-e^{-\gamma u} \right) . \label{eq:sigma(u)}
\end{equation}
In analogy with the derivation in \cref{sec:noiseless}, we now change reference frame to $\vb{X}\equiv \vb{Z}-\expval{\vb{Z}}_\T{ss}$ in \cref{eq:effective_equation}, we take the average over thermal fluctuations, and we linearize the resulting equation. This way we find the evolution of the \textit{average} position $\expval{\vb{X}(t)}$ to be given by
\begin{align}
    \partial_t \expval{X_j(t)} =& -\expval{X_j(t)} \left[ \gamma+ \hat \Gamma_j(s=0) \right] \n\\
    &+ \int_{-\infty}^t \dd{u} \Gamma_j(t-u)\expval{X_j(u)} ,
    \label{eq:new_linearized}
\end{align}
which is formally the same as \cref{eq:linearized_compact}, but where $X_j(t)$ is replaced by $\expval{X_j(t)}$, the initial time $t_0$ is set to $-\infty$, and the memory kernel 
is replaced by
\begin{align}
    \Gamma_j(t) \equiv \lambda^2 \nu \int & \dslash{q} q_j^2 |V_q|^2 e^{-q^2 \sigma^2(t)} \n\\
    &\times \left[ \chi_q(t)+\nu q^2 e^{-\gamma t} C_q(t)  \right].
    \label{eq:new_memory_kernel}
\end{align}
As expected, compared to the memory kernel for the noiseless case in \cref{eq:memory_kernel} \rev{---} which includes only the first term on the r.h.s.~of \cref{eq:new_memory_kernel} \rev{---} the present one \rev{involves} a second term $\propto C_q(t)$ due to thermal fluctuations.
Note that the integration in the variable $u$ in \cref{eq:new_linearized} runs from $-\infty$, and this fact prevents a direct solution of the equation of motion by using the Laplace transform \cite{Di_Terlizzi_2020}. 
However, in order to determine the response of the average particle position to a sudden displacement $\vb{X}_0$ imposed at time $t=0$ from its stationary value $\expval{\vb{X}}=0$, one can look for a solution $\expval{\vb{X}(t)}$ of \cref{eq:new_memory_kernel} 
with $\expval{\vb{X}(t)}\equiv 0$ for $t<0$, and $\expval{\vb{X}(t)}=\vb{X}_0$ at $t=0$. 
In this way, $\expval{\vb{X}(t)}$ for $t>0$ follows immediately from a Laplace transform as in \cref{eq:X(s)}, with $\expval*{\hat X_j(s)}$ in place of $\hat X_j(s)$, and with the memory kernel $\Gamma_j(t)$ given by the 
new expression in \cref{eq:new_memory_kernel}.
In this case, an expression of the function $\hat \Gamma_j(s)$ in closed form (such as the one found in \cref{sec:modelA} in the noiseless limit) cannot be obtained. 
In spite of this complication, 
studying the strong-confinement limit 
provides already valuable information concerning the main effects of thermal fluctuations. In fact, in 
\rev{\cref{sec:discussion}}
it was shown that this limit actually captures the particle evolution
for times $t>\tau_\kappa$. Proceeding as in \cref{eq:strong_confinement}, we then inspect the formal limit 
$\gamma \to \infty$, which has the effect of suppressing 
the term proportional to the field correlator $C_q(t)$ in the memory kernel given in \cref{eq:new_memory_kernel}. Accordingly, this results into 
\begin{equation}
    \expval*{\hat X\sc(s)} =  \frac{X_0}{1 -g\left[ f(s)-f(0) \right]},
\end{equation}
where the function $f(s)$ is the same as in \cref{eq:scaling_memory} upon replacing
\begin{equation}
    V_q \mapsto \widetilde{V}_q \equiv V_q \exp[ - T q^2/(2\kappa) ].
    \label{eq:Vq-tilde}
\end{equation}
Since the role of $V_q$ is essentially that of providing a large-momentum cutoff for $q\gtrsim 1/R$ \cite{Venturelli_2022,Venturelli_2022_2parts}, we conclude that $\widetilde{V}_q$ represents an effective renormalization of the particle radius $R$, which is replaced by a combination of $R$ and the \textit{thermal length} 
\begin{equation}
    l=\sqrt{T/(2\kappa)}
    \label{eq:l-thermal}
\end{equation}
appearing in \cref{eq:Vq-tilde}. Note that $l$ coincides with the mean squared displacement of the particle in its harmonic trap due solely to thermal fluctuations. For instance, a choice of $V_q$ as in \cref{eq:potential_peak,eq:potential_peak_fourier} yields $|V_q|^2 \simeq 1-2q^2R^2$ for small $q$, so that $|\widetilde{V}_q|^2 \simeq 1-2q^2 (R^2+l^2)$, and therefore $R$ is effectively renormalized as $R \mapsto \sqrt{R^2+l^2}$.

\begin{figure*}
\centering
\subfloat[]{
  \centering
  \includegraphics[width=\columnwidth]{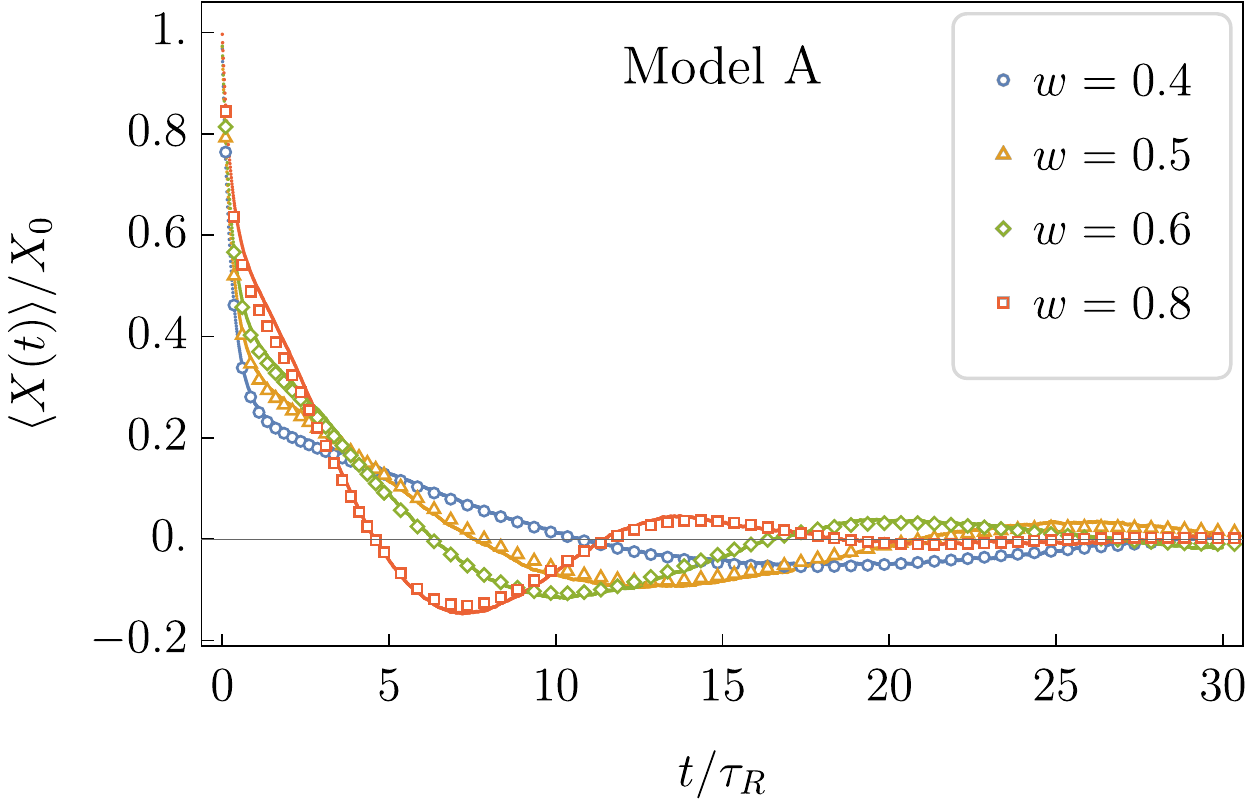}
  \label{fig:simulation_A}
  }
\subfloat[]{
  \centering
  \includegraphics[width=\columnwidth]{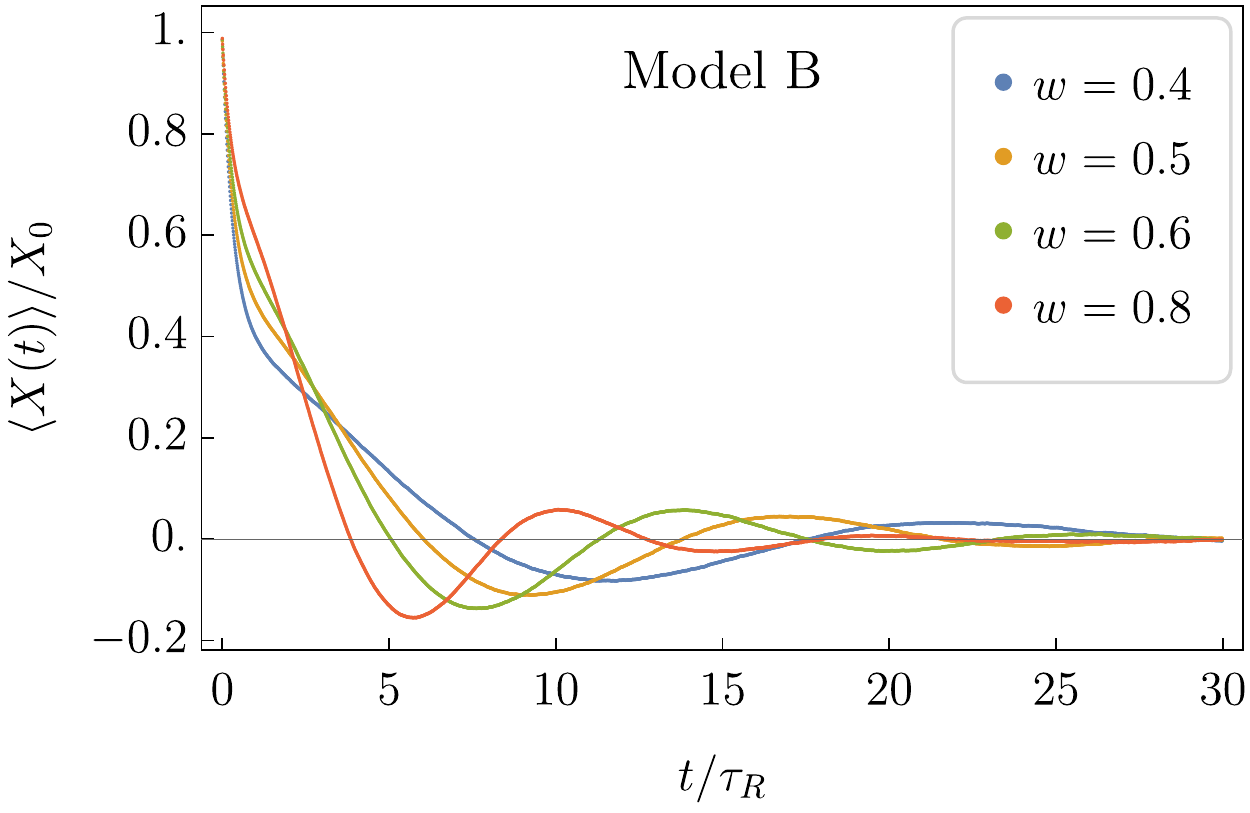}
  \label{fig:simulation_B}
  }
\caption{Evolution of the average position $\expval{X(t)}$ of the particle, after a small displacement $X_0$ from its position in the steady state --- see \cref{fig:setup}. \textbf{(a)} Critical model A in spatial dimension $d=1$. The data obtained from numerical simulations (solid lines, see the main text) are compared with the analytical prediction (symbols) calculated via the numerical inversion of the Laplace transform in \cref{eq:X(s)_rephrased,eq:modelA_1d_critical}, showing good agreement. In the simulations, the interaction potential $V_q$ was chosen to be exponential as in \cref{eq:potential_peak_fourier}. We used the parameters $\lambda=5$, $L=2500$, $r=0$, $R=5$, $\nu=5$, $\kappa=0.2$, $D=25$, and $v\in [4,8]$, corresponding to $g=25$, $\rho=1$, and $w$ as indicated in the legend. 
\textbf{(b)} Non-critical model B in spatial dimension $d=1$. In this case an analytical prediction in closed form is not available, and therefore we report only the curves obtained from numerical simulations: the qualitative behavior of the resulting evolution is similar to that of model A in panel (a). In particular, the frequency $\Omega$ of the damped oscillations increases upon increasing $w$. In this simulation the interaction potential was chosen to be Gaussian, \ie, $V_q=\exp(-q^2 R^2/2)$. We also set $R$, $\nu$, $\kappa$, and $D$ to unity, while we chose $\lambda=6$, $L=1024$, $r=0.25$, and $v\in [0.8,1.6]$, corresponding to $g=36$ with $w$ and $\rho$ as in panel (a).
In both panels (a) and (b) we chose $X_0=1$, $T=10^{-2}$, an integration timestep $\Delta t =10^{-2}$, and we averaged over $10^4$ realizations of the dynamics.
}
\label{fig:simulations}
\end{figure*}

\subsection{Numerical simulations}
\label{sec:simulation}

In this section we present and discuss the results of numerical simulations of the system in one spatial dimension, which confirm our analytical predictions, also beyond the noiseless limit presented in \cref{sec:noiseless} and
the perturbation theory discussed in \cref{sec:thermal}.
In particular, the numerical data are obtained via a direct integration of the Langevin equations \eqref{eq:particle} and \eqref{eq:field} for the particle and the field, respectively. The latter is evaluated by discretizing the field over a regular lattice with spacing $a \ll R$, similarly to \ccite{Venturelli_2022,Venturelli_2022_2parts,code}. The coupled stochastic differential equations are then integrated by using a stochastic Runge-Kutta algorithm, as described in \ccite{Roberts_2012} (which is suited for investigating also cases with an explicitly time-dependent external drag).

The field is initially prepared, at time $t=-\cor{T}$, in the flat configuration $\phi(\vb{x},t=-\cor{T}) = 0$; the harmonic potential which traps the particle is dragged for a certain time $\cor{T}$ until the system reaches its steady state, in which the average particle position stops evolving in the comoving frame of reference. 
At time $t=0$, the particle coordinate is suddenly displaced by an amount $X_0$ and its relaxation is recorded. 
Since the actual position of the particle at time $t=0^-$ depends on the realization of the noise, it fluctuates. Accordingly, the result of this displacement is equivalent to extracting the initial particle position at time $t=0^+$ from a distribution \rev{that} is the same as the one in the steady state, but shifted in space by an amount $X_0$.
We repeat the whole process (including thermalization) several times, and we finally take the average over the various realizations.
Simulations are performed with periodic boundary conditions in order to approximate the behavior of the particle in the bulk. The lattice extension $L$ is chosen sufficiently large so as to avoid \textit{stirring} effects: in fact, a particle dragged along a ring of finite length $L$ soon generates spurious field currents, which in general modify the particle statistics. The value of the particle displacement $X_0$ is chosen within the linear-response regime, which is verified a posteriori by comparing simulations performed for various (small) values of $X_0$, checking that the corresponding average particle trajectories $\expval{X(t)}$ collapse onto each other upon rescaling their amplitude by $X_0$.

Figure~\ref{fig:simulations} presents the results of the numerical simulations described above. In particular, \cref{fig:simulation_A} corresponds to the case of critical model A, which we studied analytically in \cref{sec:modelA}. For various values of the the drag velocity $v$ (which determines the value of the Weissenberg number $w$ indicated in the plot, see \cref{eq:weissenberg}), we plot $\expval{X(t)}$ (solid line) 
of a particle \rev{that} is initially displaced from its steady-state position by an amount $X_0$, as a function of the time $t$ elapsed from the displacement.
These numerical curves are compared with our analytical prediction (symbols), which is obtained by numerical Laplace inversion of \cref{eq:X(s)_rephrased,eq:modelA_1d_critical}, and in which the particle radius $R$ is replaced by the effective radius $(R^2+l^2)^{1/2}$ to account for thermal fluctuations (see discussion at the end of \cref{sec:thermal_derivation}). 
The plots show an overall agreement within the entire time range, including the fast initial decay displayed at short times. In general, this decay develops over a timescale $t\sim \tau_\kappa$, followed by an oscillating behavior which persists over a few tens of $\tau_R$. Following our discussion 
\rev{in \cref{sec:discussion}},
the latter region 
$t>\tau_\kappa$ 
in \cref{fig:simulation_A}
is essentially described by the strong-confinement limit.
This limit turns out to describe accurately the numerical data even when 
the choice of parameters is not strictly into the strong-confinement regime $\rho \gg 1$, as shown in \cref{fig:simulation_A} (see caption), which corresponds to $\tau_\kappa=\gamma^{-1} =1$, $\tau_R=1$, and therefore $\rho=1$.
This fact confirms the expectation that the phenomenology described by the dynamical phase diagram presented in \cref{fig:analytic-structure} 
actually carries over moderately beyond 
the strong-confinement limit.

Our previous discussion in \cref{sec:noiseless} revealed that the behavior of the \textit{noiseless} model is completely determined by fixing
the dimensionless numbers $g= \lambda^2/(\kappa R)$, $\rho = \gamma R^2/D$, and $w=Rv/(2D)$ --- see \cref{eq:dimensionless_coupling,eq:weissenberg,eq:rho}, here specialized for model A in spatial dimension $d=1$.
Thermal fluctuations are, instead, perturbatively quantified by the ratio $l/R$ of the thermal length $l$ (see \cref{eq:l-thermal}) to the particle radius $R$ --- see \cref{sec:thermal_derivation}. 
Note that the effective particle dynamics at criticality $r=0$ has been written 
in the previous Sections 
in terms of $n=10$ physical variables (\ie, $X$, $t$, $\kappa$, $R$, $\nu$, $\lambda$, $D$, $v$, $X_0$, and $T$), but only $k=4$ distinct physical units (i.e., mass, length, time and temperature). 
The physics of the model is thus actually captured by the mutual dependence of the $n-k=6$ dimensionless parameters
\begin{equation}
    \frac{X}{X_0} = F\left(\frac{t}{\tau_R};w,\rho,g,\frac{l}{R} \right),
\end{equation}
as suggested by 
dimensional analysis \cite{barenblatt}.
In the simulations, we chose $\rho=1$ and a large effective coupling $g\simeq 25$, while we varied the Weissenberg number $w$ within the range $0< w \lesssim 1$. 
We finally added thermal fluctuations of moderate strength by tuning the noise temperature $T$ so that $l/R\simeq 10^{-1}-10^{-2}$. This choice facilitates the numerical computation,
and we detected 
no significant qualitative change 
at higher temperatures.

\begin{figure}[t]
    \centering \includegraphics[width=\columnwidth]{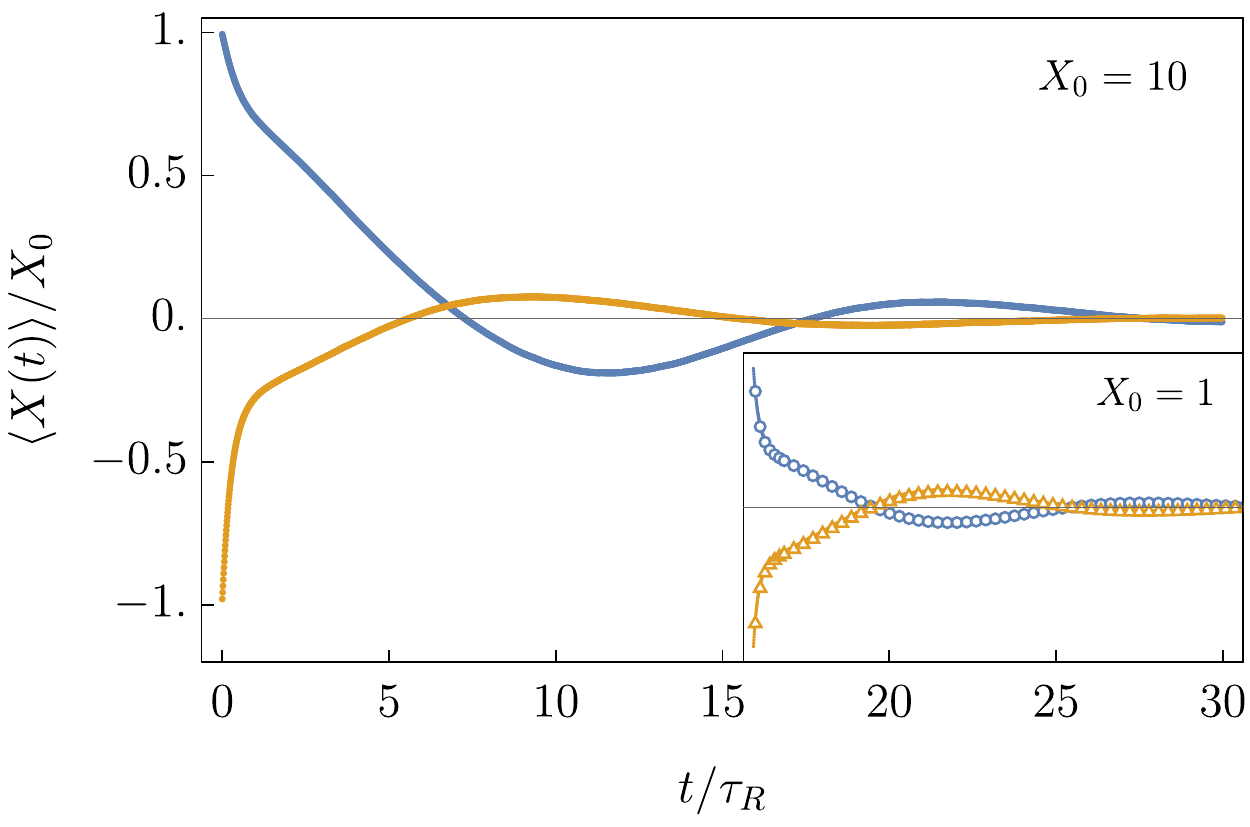}
    \caption{Evolution of the average position $\expval{X(t)}$ of the particle which is suddenly displaced at time $t=0$ by an amount $\pm X_0$ from its actual position in the steady state. The solid lines in both the main panel and the inset are obtained from the numerical simulation of the system with model A dynamics in spatial dimension $d=1$. The inset refers to the case $X_0=1$, so that the corresponding behavior is captured by the linear-response prediction in \cref{eq:X(s)_rephrased,eq:modelA_1d_critical} --- the symbols correspond to its numerical Laplace inversion, and the scales on the axes are the same as in the main plot.
    This implies, inter alia, that $\expval{X(t)}$ starting from $+X_0$ is the opposite
    of $\expval{X(t)}$ starting from $-X_0$. The main plot, instead, refers to $X_0=10$, which turns out to be beyond the linear regime. In fact, the relaxation occurring from the initial values $+X_0$ and $-X_0$ are no longer related by the symmetry highlighted above. The remaining simulation parameters are the same as in \cref{fig:simulation_A}, with $w=0.6$. 
    }
    \label{fig:beyond}
\end{figure}

In \cref{fig:simulation_B} we show the results of simulations analogous to those presented in 
\cref{fig:simulation_A}, but for a field \rev{that} evolves according to the conserved dynamics prescribed by model B. In addition, we chose here a finite correlation length $\xi\simeq 2 R$, so that the system is off criticality.
Although analytical predictions cannot be derived in closed form for model B, the simulations show a behavior similar to that of model A. 
This is interesting in view of possible experimental investigations of the effects qualitatively predicted in this work, because off-critical model B more realistically represents, e.g., the case of colloidal particles immersed in a binary liquid mixture \cite{Paladugu_2016,Ciliberto_2017,Magazzu_2019} (still assuming that hydrodynamic effects are negligible).
Note that the parameters $w$, $\rho$ and the ratio $l/R$ used above are not far from those realistically achievable in experiments \footnote{See, e.g., \cite{Magazzu_2019,GomezSolano_2015}. The field mobility $D$ in our model can be estimated by comparison with the relaxation time predicted by model H, which describes the critical dynamics in binary liquid mixtures (see Chapter~6.1.3 in \cite{onuki} and Appendix~J in \cite{Venturelli_2022}), and assuming that the correlation length $\xi$ of the medium can be made almost comparable with the radius $R$ of the colloidal particle.}; the magnitude of $g$ depends, instead, on the specific mechanism \rev{that} couples the medium with the particle and, from our discussion, one expects the overall effect to be enhanced if $g$ can be made large in an experimental realization.
Note also that the interaction potential $V(\vb{x})$ in \cref{fig:simulation_B} is chosen to be Gaussian with variance $R$, rather than exponential as in \cref{eq:potential_peak}. The overall qualitative behavior is thus shown to be robust against changing the details of $V(\vb{x})$, as expected \cite{Venturelli_2022}.

Finally, we can use the numerical simulations to explore qualitative features \rev{that} are not captured by the linear-response analysis. In particular, our analytical solution in \cref{eq:X(s)} depends linearly
on the initial particle displacement $X_0$, meaning that the evolution of $\expval{X(t)}$ after displacing the particle in the steady state by $+X_0$ is expected to be the opposite of that of $\expval{X(t)}$ after a displacement $-X_0$. 
This is indeed the case in our numerical simulations performed at small $X_0$, as we show in the inset of \cref{fig:beyond} (it is also mostly the case in the experiments of \ccite{Berner2018} --- see Fig.~3 therein). 
However, the asymmetry between the two evolutions is expected to emerge upon increasing $X_0$, as it is clearly shown in \cref{fig:beyond}. This asymmetry is a consequence of the non-linearity of the field-particle coupling, and therefore of the effective evolution equation for the particle position.

\section{Summary and conclusions}
\label{par:conclusion}

In this work we considered an overdamped Brownian particle dragged at constant velocity by a harmonic trap through a correlated medium, modeled here by a scalar Gaussian field with an overdamped Langevin dynamics. We have demonstrated that, when displaced from its position in the steady state, the resulting average position of the particle can exhibit oscillations during relaxation, in spite of the system dynamics being overdamped. This is reminiscent of the oscillatory modes recently observed with colloidal particles dragged through a viscoelastic fluid \cite{Berner2018}, except that the medium considered in this work is not viscoelastic. 
Accordingly, we have shown that oscillating modes can be found in overdamped media characterized by spatial and temporal correlations, which is typically the case for systems close to a second-order phase transition --- such as those involved when studying defects moving within spin systems \cite{demery2010-2}, or colloidal particles immersed in binary liquid mixtures close to the critical point of their demixing transition \cite{GambassiCCF}.

In particular, in \cref{sec:noiseless} we first neglected thermal fluctuations and we derived an analytic solution of the effective equation of motion for the coordinate $X(t)$ of the particle, within the linear-response approximation, and when the particle is suddenly displaced at time $t=0$ from its steady-state position --- see \cref{eq:X(s)}. 
This approximation involves the field-induced memory kernel $\Gamma(t)$ which appears into the effective equation \eqref{eq:linearized_compact} of the particle, once the field coordinate has been integrated out.
We then focused on the case $d=1$ of model A dynamics, and we characterized the analytic structure of the memory kernel $\hat \Gamma(s)$ in Laplace space (see \cref{eq:modelA_1d_critical} and \cref{fig:analytic-structure}), along with its implications for the dynamics of $X(t)$. 
In particular, it turns out that $\expval{X(t)}$ generally exhibits oscillations if the relaxation timescale of the field $\tau_R$ (over distances of the order of the particle size $R$, see \cref{eq:tau_R}) exceeds the typical timescale $\tau_\kappa$ set by the harmonic trap (see \cref{sec:phase-diagram}). These oscillations are damped (and eventually vanish) at large values of the trap strength $\kappa$, and whenever the correlation length of the field $\xi \ll R$ (i.e., far from the critical point).

Thermal fluctuations were then reinstated into the problem by using a perturbative expansion in the field-particle coupling $\lambda$ --- see \cref{sec:thermal}. Their main effect on the late-time particle dynamics is a renormalization of the particle radius $R$ by its thermal mean squared displacement $l$ in the harmonic trap (see \cref{eq:l-thermal}), while the qualitative features of $\expval*{X(t)}$ remain the same as in the absence of the noise. 

The accuracy of our analytic predictions was tested via numerical simulations in \cref{sec:simulation}, finding good agreement (see \cref{fig:simulations}). However, simulations can also be used to explore \rev{a} range of parameters \rev{that} are in principle out of reach of our analytical predictions. 
For example, in \cref{fig:simulation_B} we showed that the qualitative features of $\expval*{X(t)}$ obtained by using a conserved  field dynamics, \ie, model B, are similar to those of model A. Moreover, such features are robust against changing the particular shape of the field-particle interaction potential $V(\vb{x})$ (see \cref{eq:Hint}), as expected. In addition, by choosing a sufficiently large value of the initial particle displacement $X_0$, we can go beyond the linear-response approximation under which our analytical predictions were derived. In \cref{fig:beyond}, the actual non-linearity of the field-particle coupling causes an asymmetry between the response of the system to a $+ X_0$ or $-X_0$ initial particle displacement.

Our work opens the possibility of observing oscillatory 
modes with colloidal particles in near-critical binary liquid mixtures, in a  fashion similar to that described in \ccite{Berner2018}. 
This type of systems is already accessible experimentally \cite{Hertlein_2008,Gambassi_2009,Paladugu_2016,Ciliberto_2017,Magazzu_2019}, and correlation lengths of the order of microns (which is the typical size of a colloidal particle) can nowadays be obtained by using, e.g., micellar solutions.
It would  be therefore desirable to test if the phenomena described in this work can also be observed in experiments, at least qualitatively. 
In fact, a more quantitative description requires to incorporate hydrodynamic effects into the model, which is left for future investigations. 

Among the various additional aspects of the dynamics of the system considered here, its stochastic thermodynamics turns out to be particularly rich \cite{heat}.
%
Moreover, we note that other interesting features displayed by viscoelastic fluids --- such as those observed in the \textit{recoil} experiments performed in \ccite{GomezSolano_2015} --- are found to emerge also within the minimal model for correlated (but not viscoelastic) media studied here.
These interesting similarities will be explored in future works.
Finally, it would be interesting
to consider the case in which the field interacting with the tracer particle is \textit{active}: this might provide a model of a nonequilibrium bath made of active particles, such as those recently investigated in \ccite{Maes_2020,Granek_2022,GuevaraValadez2023,Santra2023}.

\begin{acknowledgments}

We thank Clemens Bechinger and F\'elix Ginot for useful insights. DV would like to thank Guido Giachetti and Ignacio A.~Mart\'{\i}nez for fruitful discussions. We also thank Sarah A. M. Loos, {\'E}dgar Rold{\'a}n, and Benjamin Walter for collaboration on related topics.
AG acknowledges support from MIUR PRIN project ``Coarse-grained description for non-equilibrium systems and transport phenomena (CO-NEST)” n.~201798CZL.
\end{acknowledgments}

\appendix

\section{Dynamics of the free field}
\label{par:freefield}

The Langevin equation \eqref{eq:fieldFourier} for the field reads, for $\lambda=0$,
\begin{equation}
    \dot{\phi}_q = -\alpha_q \phi_q + \zeta_q ,
    \label{eq:free_field}
\end{equation}
where $\alpha_q$ was defined in \cref{eq:tau_phi}, and the noise correlations are given in \cref{eq:field_noise}.
This equation is the same as that for the coordinate $\phi_q$ of an Ornstein-Uhlenbeck particle \cite{risken}. Accordingly, by setting for simplicity $\phi_q(t_0)\equiv 0$ (a choice which is inconsequential in the steady state which we focus on in this work), one can easily derive \cite{Tauber}
\begin{equation}
    \expval*{\phi_q(s_1)\phi_p(s_2)}_0 = \delta^d(\vb{p}+\vb{q})C_q\z(s_1,s_2) ,
\end{equation}
where
\begin{equation}
    C_q\z(s_1,s_2) = \frac{T}{q^2+r} \left[ e^{-\alpha_q |s_2-s_1|} - e^{-\alpha_q (s_1+s_2-2t_0)} \right] 
    \label{eq:freefieldcorrelator}
\end{equation}
is the free-field correlator. By formally taking the limit $t_0\rightarrow -\infty$ in \cref{eq:freefieldcorrelator}, one obtains the equilibrium correlator 
\begin{equation}
    C_q\z(t) = \frac{T}{q^2+r} e^{-\alpha_q \abs{t}}, \label{eq:field_correlator_zero}
\end{equation}
which is a function of the time difference $t=s_2-s_1$ only. 
The response function $G_q\z(t)$ and the linear susceptibility $\chi_q\z(t)$   of the free field are usually defined \cite{Tauber} as 
\begin{align}
    \chi_q\z(t) &= Dq^\alpha G_q\z(t), \label{eq:field_susceptibility_zero} \\
    G_q\z(t) &= e^{-\alpha_qt}\Theta(t).
\end{align}
These quantities are related to the equilibrium correlator $C_q\z(t)$ in \cref{eq:field_correlator_zero} by the fluctuation-dissipation theorem
\begin{equation}
    T \chi_q\z(\tau) = -\Theta(\tau) \pdv{\tau} C_q\z(\tau),
\end{equation}
where $\Theta(\tau)$ is 
the Heaviside theta function.
%
%

\section{Behavior in the time domain from the analytic structure of the Laplace transform}
\label{app:tauberian}

The features of the long-time behavior of a function $f(t)$ can be inferred from the analytic structure of its Laplace transform $\hat f(s)$. Relations of this type are referred to as \textit{Haar's Tauberian theorems} in the mathematical literature \rev{\cite{hull_froese_1955}}. In this Appendix we recap and summarize some useful related results which are applied in \cref{sec:noiseless}.

\emph{Simple poles.} --- Consider, first of all, the textbook 
case \cite{Schiff_1999}
in which $\hat f(s)$ is a meromorphic function in the complex $s$ plane:
\begin{equation}
    \hat f(s) = g(s) \prod_{j=1}^n \frac{1}{s-s_j},
\end{equation}
where $g(s)$ is an analytic function, and the $n$ poles $\{ s_1,s_2,\ldots, s_n\}$ are located at $s_j=-\delta_j+i\Omega_j$. We order the poles so that $0\leq \delta_1<\dots<\delta_n$.
By using the Cauchy residue theorem we then easily obtain
\begin{align}
    f(t) &= \sum_{j=1}^n g(s_j) e^{-\delta_j t +i\Omega_j t} \prod_{k\neq j} \frac{1}{s_j-s_k} \n\\
    &\simeq g(s_1) \left(  \prod_{k\neq 1} \frac{1}{s_1-s_k} \right) e^{-\delta_1 t +i\Omega_1 t},
    \label{eq:dominant_pole}
\end{align}
where in the last step we retained the dominant term at long $t>0$. This shows that the rightmost pole $s_1$ of $\hat f(s)$ (i.e., the closest to the imaginary axis) determines the long-time behavior of $f(t)$, which exhibits damped oscillations with frequency $\Omega_1$ whenever $s_1$ has a nonzero imaginary part.

\emph{Branch cuts.} ---
Next, assume that $\hat f(s)$ is no longer meromorphic, but rather displays a branch cut with branch point $s_0$ (for instance, in \cref{fig:poles} the branch cut develops along the real axis for $\Re{s}< -w$, with $s_0=-w$).
In this case, we can generally expand $\hat f(s)$ around the branch point $s_0$ as
\begin{equation}
    \hat f(s) \sim \sum_j a_j (s-s_0)^{\lambda_j},
\end{equation}
for some (possibly non integer) $\lambda_j$, and
take the inverse Laplace transform term by term as in \cref{eq:bromwich} to obtain
\begin{equation}
    f(t) \sim e^{s_0 t} \sum_j \frac{a_j}{\Gamma_E (-\lambda_j) \,t^{1+ \lambda_j}}.
    \label{eq:branch_point}
\end{equation}
In the presence of poles alongside the branch cut (as in \cref{fig:poles}), the contribution in \cref{eq:branch_point} simply adds up to that in \cref{eq:dominant_pole}. Again, the long-time behavior of $f(t)$ is determined by the rightmost among the poles $s_j$ and the branching point $s_0$.

\emph{Algebraic decays.} ---
We describe for completeness the case in which $f(t)$ does not exhibit an oscillatory behavior, but rather an asymptotic algebraic decay of the form
\begin{equation}
    f(t) \sim A t^{-\mu}, \quad \mbox{for}\quad t\geq t_c \gg 1 ,
    \label{eq:asymptotic_mu}
\end{equation}
where $\mu \geq 0$, and $t_c$ is a crossover time \footnote{Simpler heuristic arguments can be found in the literature for the case $0\leq \mu <1$ --- see, e.g., Refs.~\cite{Redner_2001,Margado_2002}}. The strategy to obtain the corresponding Laplace transform is to divide the integration domain as
\begin{equation}
    \hat f(s) = \int_0^{t_c} \dd{t} e^{-st} f(t) + A \int_{t_c}^\infty \dd{t} e^{-st} t^{-\mu} .
    \label{eq:first_split}
\end{equation}
The first term is regular, i.e., it can be expanded in a power series containing only integer powers of $s$. The integration in the second term of \cref{eq:first_split} can be further split into \footnote{For the sake of the argument we are assuming here $s\in \mathbb{R}$, but the resulting series expansion in \cref{eq:laplace_expansion} is well defined on a compact region of the real axis, and it can thus be analytically continued to all $s\in\mathbb{C}$.}
\begin{align}
    &\int_{t_c}^\infty \dd{t} e^{-st} t^{-\mu} = \int_{t_c}^{1/s} \dd{t} e^{-st} t^{-\mu} + \int_{1/s}^\infty \dd{t} e^{-st} t^{-\mu} \n \\
    &=\int_{t_c}^{1/s} \dd{t} e^{-st} t^{-\mu} + s^{\mu-1}  \int_1^\infty \dd{\tau} e^{-\tau} \tau^{z-1} .
    \label{eq:divide_et_impera}
\end{align}
The first term on the r.h.s. of \cref{eq:divide_et_impera} 
can be shown to involve both a regular and a non-regular part. To see this, we expand the exponential in power series and integrate term by term to find
\begin{equation}
    \cor{I} \equiv \int_{t_c}^{1/s} \dd{t} e^{-st} t^{-\mu} = \sum_{n=0}^\infty \frac{(-1)^n\, (s^{\mu-1}-s^n \, t_c^{n-\mu+1})}{(n-\mu+1)n!}.
    \label{eq:series_I}
\end{equation}
For $\mu$ not integer, the second term in the series above is regular. The first term $\sim s^{\mu-1}$, together with the last term in \cref{eq:divide_et_impera}, reconstructs the Euler gamma function via its integral representation (for $z\neq 0,-1,-2,\dots$ \cite{NIST_DLMF})
\begin{equation}
    \Gamma_E(z) = \int_1^\infty \dd{\tau} \tau^{z-1} e^{-\tau}+ \sum_{n=0}^\infty \frac{(-1)^n}{(n+z)n!} .
\end{equation}
If instead $\mu \in \mathbb{N}^+$, let us first set $\mu=p+\varepsilon$, with $p \in \mathbb{N}^+$ and $\varepsilon\ll 1$, and isolate the diverging term in \cref{eq:series_I} as
\begin{align}
    \cor{I} &= \sum_{n\neq (p-1)}^\infty \eval{(\dots)}_{\mu=p} - \lim_{\varepsilon\to 0} \frac{(-s)^{p-1}}{(p-1)!\, \varepsilon} \left( s^\varepsilon-t_c^{-\varepsilon} \right) \n \\
    &= \sum_{n\neq (p-1)}^\infty \eval{(\dots)}_{\mu=p} - \frac{(-s)^{p-1}}{(p-1)!\,} \ln(s\, t_c) ,
\end{align}
where we used $x^\varepsilon=e^{\varepsilon \ln x}\simeq 1+\varepsilon \ln x $. 
Note that the remaining infinite series may still produce terms proportional to $s^m \ln s$, with $m>p-1$, but these are subleading for small $s$.

Including all the terms in \cref{eq:first_split} we thus finally get
\begin{align}
    &\hat f(s) = \hat f_r(s) +    \label{eq:laplace_expansion} \\
    &+\begin{dcases}
        A \, \Gamma_E(1-\mu)\, s^{\mu-1}, & \mu \in [0,\infty) \smallsetminus \mathbb{N}^+, \\
                \frac{A(-1)^\mu}{(\mu-1)!} \ln(s\, t_c) s^{\mu-1} +\order{s^\mu \ln s }, & \mu \in \mathbb{N}^+,\\
    \end{dcases}\n
\end{align}
where $\hat f_r(s)$ is the regular part of $\hat f(s)$. Thus, to unveil a long-time asymptotic power-law decay of $f(t)$ \rev{as in \cref{eq:asymptotic_mu}}, one can expand its Laplace transform $\hat f(s)$ in series for small $s$, and check for the presence of a term $\sim \ln(s t_c) s^{\mu-1}$ (with $\mu$ a positive integer), or $\sim s^{\mu-1}$ (with $\mu\geq 0$ not integer).
An application of these last relations is presented in \cref{app:relaxation}.

\section{Relaxation towards equilibrium}
\label{app:relaxation}

We consider here the problem of the relaxation towards equilibrium of a particle in contact with a scalar Gaussian field, in a \textit{fixed} harmonic trap, and which is subject to a small initial displacement $\vb{X}_0\neq 0$ at time $t=0$.
Indeed, in the absence of external dragging the steady state reached by the system at long times (see \cref{sec:steady state}) is actually an equilibrium state \cite{Venturelli_2022}.

The problem is analogous to the one we analyzed in \cref{sec:noiseless} upon setting $v=0$, and thus the solution $\hat X_j (s)$ for $T=0$ (noiseless case), and within the linear-response approximation, is given by \cref{eq:X(s)}. In particular, the memory kernel in \cref{eq:memory_kernel_scaling,eq:scaling_memory} reduces to
\begin{equation}
    \hat \Gamma_j (s) =  \lambda^2 \nu D \int \dslash{q} \frac{q_j^2 q^\alpha |V_q|^2}{s+\alpha_q}.
    \label{eq:Gamma_v0}
\end{equation}
At the critical point $r=0$ of the medium one has $\alpha_q=Dq^z$ (see \cref{eq:tau_phi}), so that, using polar coordinates and changing variables to $y=Dq^z/s$, one finds
\begin{equation}
\label{eq:Gamma_relaxation}
    \hat \Gamma_j (s) = \lambda^2 \nu c_d (s D)^{d/z} \int_0^\infty \frac{\dd{y} y^{d/z}}{1+y} |V_{(sy/D)^{1/z}}|^2 \sim s^{d/z}.
\end{equation}
Here $c_d$ is a numerical constant accounting for the integration over the angular variables, while in the last step we expanded the expression for small $s$ by using the normalization condition $V_q=1+\order{q}$ of the interaction potential. 
Expanding the denominator of \cref{eq:X(s)} in a geometric series now gives
\begin{equation}
    \hat X_j(s) = X_0 \sum_{n=0}^\infty  \left[ \hat \Gamma_j(s) -s  \right]^n    \left[ \hat \Gamma_j(0) +\gamma  \right]^{-(n+1)}  .
    \label{eq:X(s)_geometric}
\end{equation}
Comparing with \cref{eq:Gamma_relaxation}, we deduce that the power series of $\hat X_j(s)$ contains a term $\sim s^{d/z}$, which is non-regular whenever the ratio $d/z$ is not integer. From our discussion in \cref{app:tauberian}, this corresponds to an algebraic asymptotic decay of $X_j(t)\sim t^{-\mu}$, with $\mu=1+d/z$. Note that the terms $\sim s^{nd/z}$ contained in the series of \cref{eq:X(s)_geometric} may also be non-regular, but they correspond to subleading algebraic contributions to $X_j(t)$ at long times.

Similarly, for critical model B one has $\alpha_q=Dq^2(q^2+r)$ (see \cref{eq:tau_phi}). Accordingly, by following the same steps as those \rev{that} led from \cref{eq:Gamma_v0} to \cref{eq:Gamma_relaxation}, one finds
\begin{equation}
    \hat \Gamma_j (s) = \lambda^2 \nu c_d  \int_0^\infty \frac{\dd{y} |V_{(sy/D)^{1/2}}|^2}{1+y(r+sy/D)}  \left(\frac{sy}{D}\right)^{1+d/2} \sim s^{1+d/2} ,
\end{equation}
where we set $y\equiv Dq^2/s$.
Comparing with \cref{eq:X(s)_geometric} one eventually concludes that $X_j(t)\sim t^{-\mu}$ with the decay exponent $\mu=2+d/2$.

The powers of the algebraic decays of $X_j(t)$ agree with those previously found in \ccite{Venturelli_2022} --- see Eqs.~(33) and (34) therein. The approach used in Ref.~\cite{Venturelli_2022} in order to derive these predictions differs, however, from the one used here and in \cref{sec:noiseless}: in particular, the former relies on a weak-coupling expansion for small $\lambda$, while it assumes $T\neq 0$ and it does not require the linear-response approximation (thus it also allows the investigation of an intermediate nonlinear dynamical crossover, see Ref.~\cite{Venturelli_2022} for details). 
Although limited to the noiseless case $T = 0$, the analysis presented here and in \cref{sec:noiseless} of the present work makes, instead, no assumption concerning the magnitude of $\lambda$. This suggests that the exponents of the algebraic decays determined above may in fact be nonperturbative in $\lambda$, as was conjectured in Ref.~\cite{Venturelli_2022} based on the evidence provided by  numerical simulations.
%

\begin{figure*}
\centering
\subfloat[]{
  \centering
  \includegraphics[width=\columnwidth]{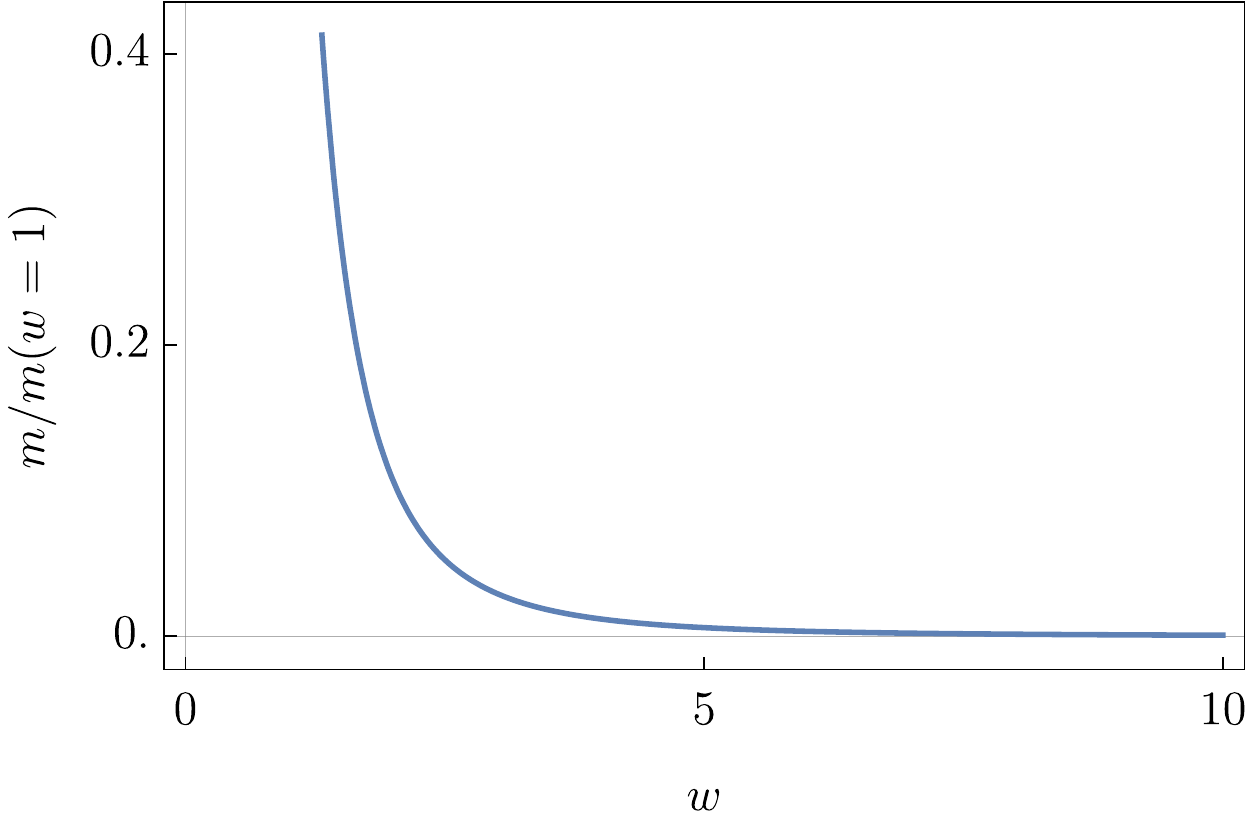}
  \label{fig:effective-mass}
  }
\subfloat[]{
  \centering
  \includegraphics[width=\columnwidth]{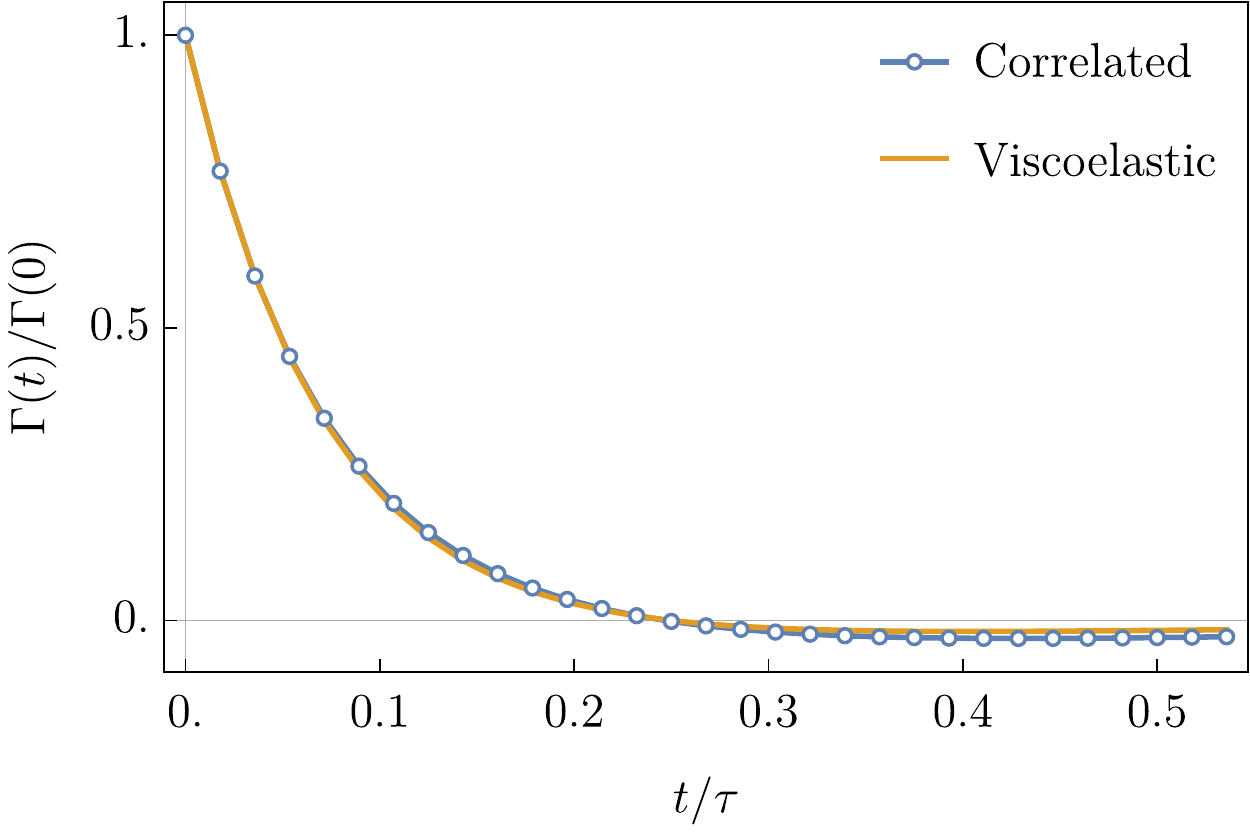}
  \label{fig:kernels}
  }
\caption{Comparison with the phenomenological model for viscoelastic fluids of Ref.~\cite{Berner2018}. \textbf{(a)} Effective mass $m$ given in \cref{eq:effective_mass} as a function of $w$ (see the main text for details). \textbf{(b)} Comparison between the memory kernel $\Gamma (t)$ of a viscoelastic fluid (solid yellow line, obtained as $\Gamma(t) = -\nu \cor{K}'(t)$ from \cref{eq:kernel_jeffrey}) and the one of the model studied in this work (blue line with symbols, corresponding to \cref{eq:kernel_modelA_time} for $d=1$ model A). 
    The parameters of the former kernel were chosen as in Ref.~\cite{Berner2018} (see the main text). For illustrative purposes, in the latter kernel in \cref{eq:kernel_modelA_time} we set $\tau_\xi=\infty$ (\ie, we considered the field at criticality), while we fixed the parameters $\tau_R$ and $\tau_v$ by matching the leading behavior of the former $\Gamma(t)$ at short times, and the crossing point of the horizontal axis. Both curves eventually approach zero from below as $t\to\infty$.}
\end{figure*}

\section{Comparison with a phenomenological model for viscoelastic fluids}
\label{app:berner}

The underdamped oscillations of a colloidal particle dragged through a viscoelastic fluid reported in \ccite{Berner2018} have been described therein in terms of the linear, generalized Langevin equation
\begin{equation}
    \gamma_\infty \dot{X}(t) + \int_{-\infty}^t \dd{u}  \cor{K} (t-u) \dot{X}(u) = -\kappa X(t) + f\v(t) ,
    \label{eq:GLE_Berner}
\end{equation}
where $\gamma_\infty$ is the friction coefficient at infinite frequency, while $f\v(t)$ is a stochastic correlated  noise term with vanishing average.
\rev{As in \cref{sec:noiseless}, $X(t)$ denotes the component of the particle position along the direction of the trap displacement, in the comoving frame of reference.}
Using that both $\expval*{X(t)}$ and $\expval*{\dot{X}(t)}$ vanish for $t<0$ and integrating by parts, one can check that \cref{eq:GLE_Berner} is formally equivalent to the effective equation in \eqref{eq:new_linearized} upon identifying 
\begin{equation}
    \gamma_\infty \equiv 1/\nu , \quad \mbox{and}\quad \Gamma(t) \equiv -\nu \cor{K}'(t).
    \label{eq:mapping}
\end{equation}
Finally, the memory kernel is assumed in \ccite{Berner2018}
to be of the form 
\begin{equation}
    \cor{K}(t) = \frac{\gamma_0-\gamma_\infty}{\tau } e^{-t/\tau} + \sum_i \left( \frac{\gamma_i}{\tau_i} e^{-t/\tau_i} - \frac{\gamma_i}{\tau} e^{-t/\tau} \right),
    \label{eq:kernel_jeffrey}
\end{equation}
where $\gamma_0 > \gamma_\infty$ is the \textit{zero-frequency} friction coefficient,
while $\gamma_i$ and $\tau_i$ are phenomenological parameters introduced to fit the experimental data.
The kernel in \cref{eq:kernel_jeffrey}
reduces to that of a Jeffrey's fluid \cite{Raikher_2013} for $\gamma_i\equiv 0$, which is shown in \ccite{Berner2018} to appropriately describe the particle dynamics in a static trap, \ie, for $v=0$. For $v>0$, instead, one can rationalize the experimental data by considering one or more additional relaxation timescales $\tau_i>\tau$, weighted as in \cref{eq:kernel_jeffrey} by some suitable coefficients $\gamma_i<0$. 
It actually turns out that an increasing number of pairs $(\gamma_i,\tau_i)$ is needed for fitting the experimental data upon increasing the dragging velocity $v$, and hence the Weissenberg number \cite{Elephant_2010}.

Integrating by parts the second term on the l.h.s.~of \cref{eq:GLE_Berner} and introducing
$\cor{M}(t) \equiv -\int_t^\infty \dd{u} \cor{K}(u)$,
one can cast
\cref{eq:GLE_Berner} 
in the form
\begin{equation}
    \int_{-\infty}^t \dd{u} \cor{M} (t-u) \ddot{X}(u) =-\gamma_0 \dot{X}(t) -\kappa X(t) + f\v(t) .
    \label{eq:GLE_Berner_2}
\end{equation}
At long times, the l.h.s.~may be further approximated as
\begin{align}
&\int_{-\infty}^t \dd{u_0} \cor{M} (t-u_0) \ddot{X}(u_0)     \label{eq:berner_approx} \\
&= \int^\infty_0 \dd{u} \cor{M} (u) \left[ \ddot{X}(t) -u\dv[3]{X(t)}{t} +\frac{u^2}{2}\dv[4]{X(t)}{t}+\dots \right] \n \\
    &= \hat{\cor{M}}(0) \ddot{X}(t) + \hat{\cor{M}}'(0) \dv[3]{X(t)}{t} + \frac{1}{2} \hat{\cor{M}}''(0) \dv[4]{X(t)}{t} +\dots \n
\end{align}
where 
\begin{equation}
    \hat{\cor{M}}(s)=[ \hat{\cor{K}}(s)-\hat{\cor{K}}(0)]/s
    \label{eq:mass-laplace}
\end{equation}
denotes as usual the Laplace transform of $\cor M (t)$.
By retaining only the first term $\hat{\cor{M}}(0) \ddot{X}(t)$ in the expansion of \cref{eq:berner_approx}, then
\cref{eq:GLE_Berner_2} reduces to that of an underdamped harmonic oscillator with mass
\begin{equation}
    m \equiv \hat{\cor{M}}(0) = - \left( \gamma_0 -\gamma_\infty - \sum_i \gamma_i   \right) \tau - \sum_i \gamma_i \tau_i .
    \label{eq:effective_mass_berner}
\end{equation}
One normally finds $m<0$ when $\gamma_i\equiv 0$ (corresponding to exponentially decaying solutions for $\expval*{X(t)}$), while an appropriate choice of the coefficients $\gamma_i<0 $ can render $m>0$, \ie, a bona-fide inertia which may explain the emergence of oscillations within the system.
An oscillating behavior of $\expval*{X(t)}$ with frequency 
\begin{equation}
    \Omega = \frac{1}{2m} \sqrt{4m\kappa-\gamma_0^2} > 0
    \label{eq:harmonic_freq}
\end{equation}
is then expected for $\gamma_0 < 2\sqrt{m \kappa}$.

It is interesting to check if the analogy with a harmonic oscillator
holds for our model as well, via the mapping 
in \cref{eq:mapping}.
Considering for instance critical model A (see \cref{sec:modelA}) and using \cref{eq:mass-laplace,eq:modelA_1d_critical}, one formally finds a (zero-frequency) mass
\begin{equation}
    m =  \hat{\cor{M}}(0) = \frac{\lambda^2 \tau_R^2}{8R} \left[ \frac{1}{w^3} - \frac{2(5+4w)}{(1+2w)^4}  \right],
    \label{eq:effective_mass}
\end{equation}
as a function of the Weissenberg number $w$ (see \cref{eq:weissenberg}).
This effective mass is plotted in 
\cref{fig:effective-mass}
and it appears to be always positive, while it increases and diverges upon reducing the value of $w$ towards zero.
This behavior can be rationalized by comparison with the dynamical phase diagram in the strong-confinement limit shown in \cref{fig:phase-diagram}. 
In fact, we note that $\Omega\simeq 0$ 
in the latter as soon as the complex poles appear for small values of $w$ and $g$,
after which $\Omega$ is a growing function of $w$ (for any value of $g$). Inverting \cref{eq:harmonic_freq} yields in fact, consistently, $m \sim 1/\Omega^2$ --- \ie, heavier objects oscillate more slowly. 

It is now tempting to use the condition stated in \cref{eq:harmonic_freq} in order to predict the boundaries within the phase diagram in \cref{fig:phase-diagram}. In the strong-confinement limit of critical model A, however, the argument outlined above renders a friction coefficient $\gamma_0$ in \cref{eq:GLE_Berner_2} equal to
\begin{equation}
    \gamma_0 \to - \frac{\lambda^2 \tau_R}{4R} \frac{3+2w}{(1+2w)^3},
\end{equation}
which is negative for all values of $w$. Accordingly, this indicates that the approximation used in \cref{eq:effective_mass_berner} --- corresponding to keeping only the first order term in the expansion for small $s$ in \cref{eq:berner_approx} --- is no longer accurate in our case. 
In fact, it turns out that the nontrivial analytic structure of the memory kernel
in \cref{eq:modelA_1d_critical} 
(see \cref{fig:analytic-structure}) prevents us from simply expanding $\hat \Gamma(s)$ (and hence $ \hat{\cor{M}}(s)$) in a Taylor series around $s=0$. In order to check this, one can numerically invert $\hat X(s)$ in \cref{eq:X(s)} after replacing $\hat\Gamma(s)$ by its $n$-th order Taylor expansion: the amplitude of the corresponding approximation to $X(t)$ turns out to diverge upon increasing $t$ (unlike the actual solution, which is expected to be bounded).

In conclusion, contrary to the phenomenological model presented in \ccite{Berner2018} --- which explains the origin of the observed oscillations in terms of the emergence of an effective harmonic oscillator --- the dynamics investigated here does not admit such a simplified explanation. 
Still, it is interesting to compare the qualitative features of the memory kernels $\Gamma(t)$ \rev{that} emerge in these two cases. 
To be concrete, we consider a field with model A dynamics in $d=1$ and we choose a Gaussian interaction potential $V_q=\exp(q^2 R^2/2)$, so that the integral in \cref{eq:memory_kernel} can be computed in closed form, yielding
\begin{align}
    \Gamma(t) = & \frac{\lambda^2 \nu}{4\sqrt{\pi}R\, \tau_R} \frac{1+t/\tau_R-\frac12 (t/\tau_v)^2}{(1+t/\tau_R)^{5/2}} \n \\ 
    &\quad\times \exp[ 
-\frac{t}{\tau_\xi} -\frac{(t/\tau_v)^2}{4(1+t/\tau_R)} ].
    \label{eq:kernel_modelA_time}
\end{align}
We note that this function, which is positive for $t=0$, 
becomes negative upon increasing $t$ and, for $t\to\infty$, it approaches zero from below, provided that the dragging speed $v$ does not vanish (hence $\tau_v<\infty$, see \cref{eq:tau_v}). 

The kernel $\Gamma(t)$ in \cref{eq:kernel_modelA_time} is plotted in \cref{fig:kernels}, where we compare it to the one of \ccite{Berner2018}. The latter,
which encodes
the interaction with the viscoelastic fluid, can 
be readily obtained by combining \cref{eq:kernel_jeffrey,eq:mapping}. 
As we show in \cref{fig:kernels}, 
choices of the values of the parameters $\gamma_i$ and $\tau_i$ exist such that this second $\Gamma(t)$ also becomes negative for sufficiently large $t$. In particular, in \cref{fig:kernels} we used two 
timescales $\tau_i$, with $i\in\{1, 2\}$, as reported in Tab.~1 of \ccite{Berner2018} for $\T{Wi}=0.17$.
We thus conclude that, in both models, the memory kernel $\Gamma(t)$ features \textit{anti}-correlations at long times. 
This is reminiscent of the \textit{negative memory} often found in the context of rheology of complex fluids \cite{Sollich_98, Fielding_2000, Fuchs_2002, Falk_2011, Amann_2013}, and suggests that the underdamped modes displayed by the particle are indeed due to the negative response of the surrounding non-equilibrium environment \cite{Berner2018}, independently of its actual physical origin (\ie, due to either correlations or viscoelasticity).

\section{Correction to the steady-state particle position due to thermal fluctuations}
\label{app:correction}
\rev{Here we compute perturbatively, up to the lowest nontrivial order in the coupling constant $\lambda$, the correction to the average particle position $\expval{\vb{Z}}$ due to thermal fluctuations, \rev{discussed in Sec.~\ref{sec:thermal}}. }
\rev{We can determine $\expval{\vb{Z}}$}
by first taking the average of \cref{eq:effective_equation}, which gives
\begin{align}
    &\partial_t \expval{\vb{Z}} = -\vb{v} -\gamma \expval{\vb{Z}}  +\nu \lambda \int \dslash{q} i \vb{q} V_{-q}\times   \label{eq:effective_equation_avg}\\
    &\left[\expval{\zeta_q(t)e^{i \vb{q}\cdot \vb{Z}(t)}}+ \lambda V_q \int_{0}^\infty \dd{u} \chi_q(u) \expval{ e^{i \vb{q}\cdot \left[\vb{Z}(t)-\vb{Z}(t-u)\right]} } \right] . \n
\end{align}
The dynamical structure factor $\expval{ e^{i \vb{q}\cdot \left[\vb{Z}(t)-\vb{Z}(t-u)\right]} }$ which appears 
above was computed in \ccite{Venturelli_2022} for $\lambda=0$, finding (here we considered the formal limit $t_0\to -\infty$ of the expression in Eqs.~(A14)-(A17) 
therein)
\begin{equation}
    \expval{ e^{i \vb{q}\cdot \left[\vb{Z}(t)-\vb{Z}(s)\right]} }_0=    e^{i \vb{q}\cdot \left[\expval{\vb{Z}(t)}_0-\expval{\vb{Z}(s)}_0\right]  -q^2 \sigma^2(t-s)}, \label{eq:Qq}
\end{equation}
where we indicated by $\expval{\cdots}_0$ the average over the independent process for $\vb{Z}(t)$ (\ie, for 
$\lambda=0$), and we introduced
$\sigma(u)$ as in \cref{eq:sigma(u)}.
The expression in \cref{eq:Qq} is sufficient 
to evaluate
the r.h.s.~of \cref{eq:effective_equation_avg} at the lowest non-trivial order in $\lambda$, \ie, at $\order{\lambda^2}$. The first expectation value on the r.h.s.~of \cref{eq:effective_equation_avg}, instead, can be evaluated at $\order{\lambda}$
by using Novikov's theorem \cite{Novikov_1965,Luczka_2005}
\begin{equation}
    \expval{\zeta(t) F[\zeta]} = \int \dd{s} \expval{\zeta(t) \zeta(s)} \expval{\fdv{F[\zeta]}{\zeta(s)} } ,
\end{equation}
where $F[\zeta]$ is any functional of a Gaussian 
noise $\zeta$ --- such as the noise $\zeta_q(t)$ introduced in \cref{eq:colored_noise}, having correlations $C_q$. Accordingly, we find 
\begin{align}
    &\expval{\zeta_q(t)e^{i \vb{q}\cdot \vb{Z}(t)}} = i \vb{q} \int \dd{s} C_q(t-s) \expval{e^{i \vb{q}\cdot \vb{Z}(t)}\fdv{\vb{Z}(t)}{\zeta_{-q}(s)}  } \n\\
    & \simeq \nu \lambda q^2 V_q \int^t_{-\infty} \dd{s} e^{-\gamma (t-s)} C_q(t-s) \expval{ e^{i \vb{q}\cdot \left[\vb{Z}(t)-\vb{Z}(s)\right]} }_0,
\end{align}
where in the second line 
we used the equation of motion \eqref{eq:effective_equation} of $\vb{Z}(t)$, and we neglected higher-order terms in $\lambda$ \cite{heat,Venturelli_2022_2parts}.
Setting $\partial_t \expval{\vb{Z}}=0$ in \cref{eq:effective_equation_avg} and taking the limit $t\to \infty$ on its r.h.s.~finally yields the average position in the steady state reported in \cref{eq:Z_ss_corr},
upon using the fact
that (see \cref{eq:Qq})
\begin{equation}
    \expval{ e^{i \vb{q}\cdot \left[\vb{Z}(t)-\vb{Z}(t-u)\right]} }_0
    \xrightarrow[t\to \infty]{} e^{-q^2 \sigma^2(u)}.
\end{equation}



\bibliography{references}

\end{document}